\documentclass[Crown,sagev,times]{sagej}

\usepackage{amsmath}
\usepackage{bm}
\usepackage{bbm}
\usepackage{booktabs}
\usepackage{color}
\usepackage{enumerate}
\usepackage{enumitem}
\usepackage{float}
\usepackage{graphicx}
\usepackage{hyperref}
\usepackage{mathtools}
\usepackage{multirow}
\usepackage{natbib}
\usepackage{siunitx}
\usepackage{subcaption}
\usepackage{tcolorbox}
\usepackage{tikz}
\usepackage{twemojis}

\newcommand{\sd}{\mathrm{sd}}
\newcommand{\se}{\mathrm{se}}

\newcommand{\bbC}{\mathbb{C}}

\newcommand{\bbV}{\mathbb{V}}
\newcommand{\bbone}{\mathbbm{1}}

\newcommand{\calN}{\mathcal{N}}

\newcommand{\bfr}{\mathbf{r}}

\newcommand{\bfu}{\mathbf{u}}

\newcommand{\bfA}{\mathbf{A}}

\newcommand{\bfD}{\mathbf{D}}
\newcommand{\bfE}{\mathbf{E}}

\newcommand{\bfG}{\mathbf{G}}
\newcommand{\bfH}{\mathbf{H}}
\newcommand{\bfI}{\mathbf{I}}

\newcommand{\bfM}{\mathbf{M}}

\newcommand{\bfW}{\mathbf{W}}

\newcommand{\bfY}{\mathbf{Y}}
\newcommand{\bfZ}{\mathbf{Z}}

\newcommand{\bfbeta}{\bm{\beta}}
\newcommand{\bfepsilon}{\bm{\epsilon}}
\newcommand{\bftheta}{\bm{\theta}}

\newcommand{\sumi}{\sum_{i=1}^I}
\newcommand{\sumj}{\sum_{j=1}^J}
\newcommand{\sumq}{\sum_{q=1}^Q}

\newcommand{\sums}{\sum_{s=1}^{J-1}}

\newcommand{\newcheckmark}{\twemoji{check mark}}
\newcommand{\newcrossmark}{\twemoji{multiply}}

\begin{document}

\runninghead{}

\title{Can discrete-time analyses be trusted for stepped wedge trials with continuous recruitment?}

\author{Hao Wang$^{1, 2}$, Guangyu Tong$^{1, 2, 3}$, Heather Allore$^{1,  4}$, Kelsey L. Grantham$^5$, Monica Taljaard$^{6, 7}$, and Fan Li$^{1, 2, 3, \ast}$ \let\thefootnote\relax\footnote{
    $^1$Department of Biostatistics, Yale School of Public Health, CT, USA\\
    $^2$Center for Methods in Implementation and Prevention Science, Yale School of Public Health, CT, USA\\
    $^3$Section of Cardiovascular Medicine, Department of Internal Medicine, Yale School of Medicine, New Haven, CT, USA\\
    $^4$Section of Geriatrics Medicine, Department of Internal Medicine, Yale School of Medicine, New Haven, CT, US\\
    $^5$School of Public Health and Preventive Medicine, Monash University, Melbourne, Australia\\
    $^6$Methodological and Implementation Research Program, The Ottawa Hospital Research Institute, Ottawa, ON, Canada\\
    $^7$School of Epidemiology and Public Health, University of Ottawa, Ottawa, ON, Canada}}

\corrauth{$^{\ast}$Fan Li, Department of Biostatistics, Yale School of Public Health, 135 College Street, New Haven, CT 06510, USA.}
\email{fan.f.li@yale.edu}

\begin{abstract}
In stepped wedge cluster randomized trials (SW-CRTs), the intervention is rolled out to clusters sequentially over multiple periods. It is common practice to analyze data from SW-CRTs using linear mixed models that treat time as discrete. However, a recent systematic review found that 95.1\% of cross-sectional SW-CRTs recruit individuals continuously over time. Despite the prevalence of continuous recruitment, there has been limited guidance on how to draw model-robust inference for such designs. In this article, we conduct simulation studies to examine the performance of discrete-time linear mixed models in the case of continuous recruitment designs with a continuous outcome. Specifically, we generate outcomes under a continuous-time exponential decay correlation structure with either discrete or continuous fixed period effects. We consider constant and exposure-time-dependent intervention effects, with and without random intervention effects across clusters. We then analyze the simulated data using linear mixed models under three discrete-time working correlation structures: simple exchangeable, nested exchangeable, and discrete-time exponential decay, with robust sandwich variance estimation. Our results demonstrate that discrete-time analysis often yields negligible bias and that the robust variance estimator with the Mancl and DeRouen correction consistently achieves nominal coverage and type I error rate. However, when recruitment patterns vary systematically between control and intervention periods, discrete-time analyses give slightly biased intervention effect estimates. Finally, we illustrate these findings by reanalyzing a completed SW-CRT.
\end{abstract}


\maketitle

\section{Introduction}\label{sec:intro}

\begin{figure}[t]
    \centering
    \includegraphics[width=\linewidth]{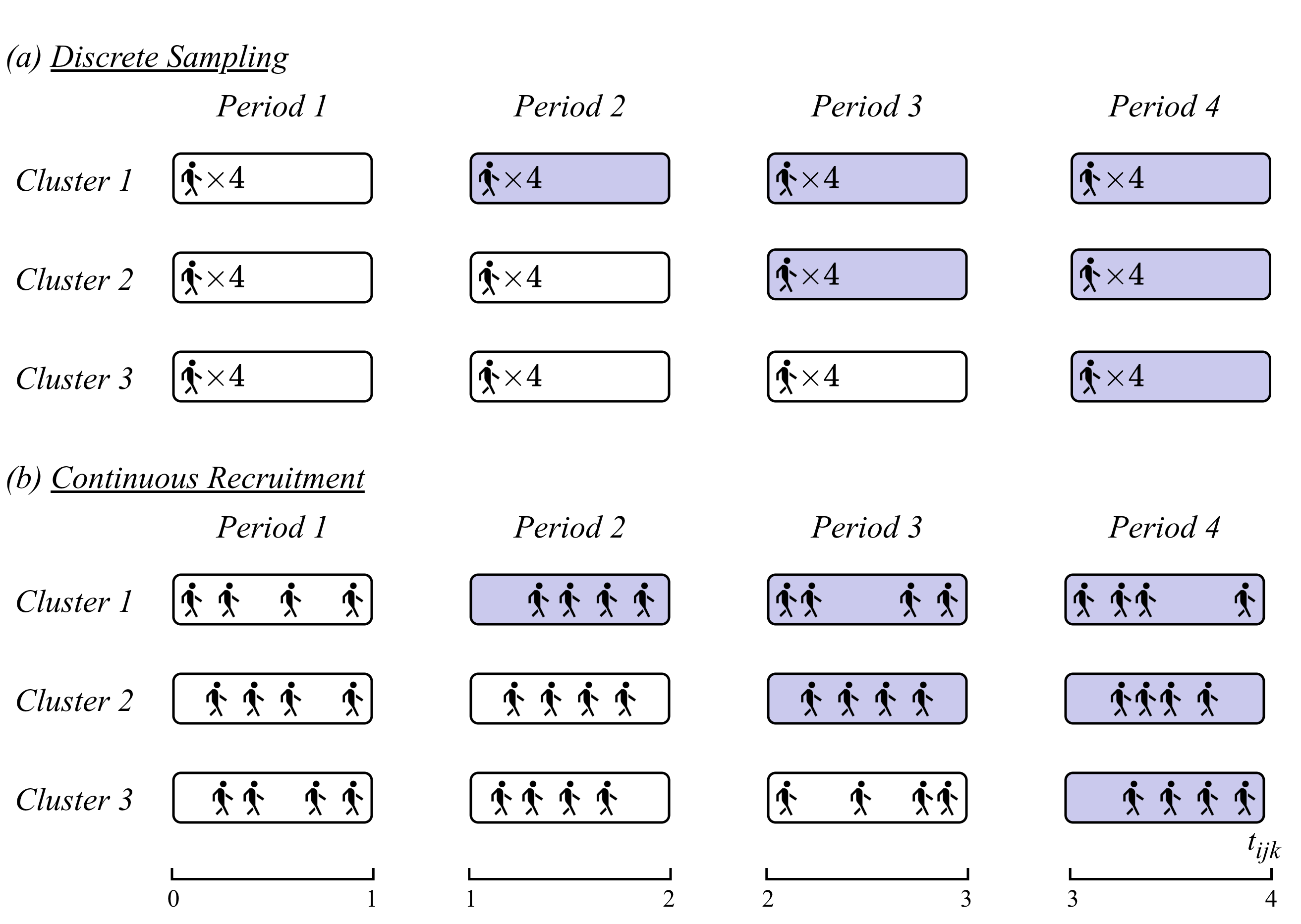}
    \caption{A comparison of two standard SW-CRTs, each consisting of 3 clusters over 4 periods of equal length, with discrete sampling in Figure~ 1(a) and continuous recruitment in Figure~ 1(b). The x-axis indicates the exact recruitment time $t_{ijk} \in (j-1, j]$ for the $k$-th individual from the $i$-th cluster in the $j$-th period. Each SW-CRT has 3 sequences with exactly one cluster assigned to each sequence, where the cluster in sequence $q \in \{1, 2, 3\}$ begins receiving the intervention in period $j = q + 1$. Purple cells indicate the intervention condition while white cells indicate the control condition. For illustration purposes, each cluster-period cell includes exactly 4 individuals. In Figure~ 1(a), all 4 individuals within each cluster-period are sampled at a single fixed time point (e.g., at the beginning of each period), representing a discrete cross-sectional slice of the cluster at that moment. In contrast, Figure~ 1(b) depicts continuous recruitment, where individuals enter the trial at distinct times within each period according to an underlying continuous-time recruitment process.}
    \label{fig:stepped_wedge_design}
\end{figure}

In stepped wedge cluster randomized trials (SW-CRTs), the intervention is rolled out to clusters (such as hospitals or clinics) sequentially over multiple periods.\cite{Hussey2007} Once a cluster crosses over to the intervention condition, it remains in that condition until the end of the trial. SW-CRTs can be cross-sectional, closed-cohort, or open-cohort, depending on whether different or the same individuals are observed in each cluster across periods.\cite{Copas2015} Of note, these categories can be viewed as points along a continuum of sampling structures, ranging from no participant overlap across periods in cross-sectional designs to complete overlap in closed-cohort designs. Open-cohort designs allow individuals to enter or leave the cohort over time, with some individuals observed in multiple periods.\cite{Kasza2020} Although these designs are often conceptualized and presented in terms of distinct time periods, in practice, individuals in cross-sectional designs are commonly recruited continuously over time. Nevins et al.\cite{Nevins2024} reviewed 160 SW-CRTs published from January 2016 to March 2022 and found that 76.3\% were cross-sectional SW-CRTs, among which 95.1\% implemented continuous-time recruitment. Nevertheless, most existing methods for cross-sectional SW-CRTs are based on regression models defined over discrete periods and rarely reflect the nature of the individual recruitment process.\cite{Hooper2019,Hooper2021} Figure~\ref{fig:stepped_wedge_design} illustrates two SW-CRTs with either discrete sampling or continuous recruitment. Under discrete sampling, individuals are recruited at fixed time points, typically once per period. In contrast, under continuous recruitment, individuals enter the trial at different time points within each period according to an underlying continuous-time recruitment process. Copas et al.\cite{Copas2015} discussed this distinction and described ``\emph{continuous recruitment short exposure designs}'' as those in which individuals are recruited continuously, exposed only briefly, and assessed once under either the control or intervention condition. Hooper and Copas\cite{Hooper2019} further emphasized that distinguishing continuous recruitment from discrete sampling is important for identifying potential contamination and informing the design and analysis of SW-CRTs.

A growing body of literature has focused on developing appropriate design methods for SW-CRTs with continuous recruitment. For instance, Grantham et al.\cite{Grantham2019} found that relying on models that treat time as discrete in the presence of continuous recruitment often leads to an underestimation of the required sample size. Hooper et al.\cite{Hooper2020} developed a computational algorithm to identify efficient incomplete designs for SW-CRTs with continuous recruitment and continuous outcomes. More recently, Hooper et al.\cite{Hooper2024} proposed optimal designs for three-sequence SW-CRTs with continuous recruitment. Despite these advances in study planning, there has been limited attention to the analysis of SW-CRTs with continuous recruitment, and it remains unclear whether discrete-time analysis of such designs delivers credible results. In this article, we tackle this practical question and investigate whether conventional discrete-time linear mixed models (i.e., linear mixed models that treat time as discrete) can be trusted when used to analyze SW-CRTs with continuous-time recruitment. The essential ingredients of the discrete-time linear mixed model include a period effect to adjust for the secular trend, the intervention effects of interest, and random effects to account for sources of heterogeneity, including within-cluster correlations.\cite{Li2020} Under continuous-time recruitment, such a linear mixed model may be misspecified. Previous studies have demonstrated that misspecification of the correlation structure under linear mixed models can lead to substantial bias in model-based variance estimators in SW-CRTs.\cite{Kasza2019b,Bowden2021,Voldal2022} More recently, Ouyang et al.\cite{Ouyang2024} illustrated through a simulation study that, with an appropriate small-sample correction, the robust variance estimator (RVE), also referred to as the sandwich variance estimator, can provide nominal confidence interval coverage for the intervention effect under linear mixed models even when the correlation structure is misspecified. Under the potential outcomes framework, Wang et al.\cite{Wang2024} proved that the linear mixed model is robust against misspecification of covariate effects, the correlation structure, and the error structure, as long as the intervention effect structure (i.e., whether the intervention effect varies by calendar time, exposure time, or both) is correctly specified. However, these theoretical results were developed under a discrete-time framework in which potential outcomes are indexed by discrete periods. Whether these robustness properties extend to the continuous-time recruitment setting, where potential outcomes are indexed by continuous recruitment times, remains an open question.

Our primary objective is to empirically investigate the implications of using discrete-time linear mixed models to analyze SW-CRTs with continuous recruitment. Three commonly used discrete-time correlation structures are the exchangeable (EXCH),\cite{Hussey2007} the nested exchangeable (NE),\cite{Girling2016,Hooper2016} and the discrete-time exponential decay (DTD)\cite{Kasza2019a,Kasza2019b} correlation structures. Additionally, random intervention effects can be included through a random cluster-by-intervention interaction term to account for treatment effect heterogeneity across clusters.\cite{Hemming2018b} For the intervention effect, different model formulations can be used depending on whether the effect is assumed to be constant or to vary with exposure time, calendar time, or both.\cite{Kenny2022,Maleyeff2022,Wang2024,Lee2025,Wang2026} Under continuous recruitment, individuals enter the trial at different times within each period, introducing a temporal ordering within each cluster-period. We therefore incorporate two additional aspects into the data-generating process to characterize continuous recruitment. First, we allow period effects to vary continuously over time. Second, we adopt the continuous-time decay (CTD) correlation structure, which assumes that correlations between individual-level outcomes within the same cluster decay exponentially as the time between measurements increases.\cite{Grantham2019,Hooper2021,Hooper2024} This decay pattern reflects the real-world scenario where individuals recruited closer in time experience more similar contextual factors. Of note, the CTD correlation structure depends on exact recruitment times and can accommodate different recruitment patterns. We represent these patterns by cluster-period-specific enrollment densities, which describe the distribution of recruitment times within each period for each cluster. Although the CTD correlation structure explicitly accounts for continuous recruitment times, fitting it as a working model can be computationally challenging or even infeasible in practice; we return to this topic in Sections \ref{sec:sim_CTD} and \ref{sec:discussion}. This practical constraint motivates our central research question: ``\emph{Can conventional discrete-time linear mixed models with robust variance estimation provide valid inference for SW-CRTs with continuous recruitment?}'' We investigate this question through extensive simulation studies and further illustrate our findings by reanalyzing a completed SW-CRT.

The rest of this article is structured as follows. In Section~\ref{sec:setup}, we introduce the setting, estimands, different linear mixed model formulations, and robust variance estimators. In Section~\ref{sec:simulation}, we present a simulation study to examine the performance of discrete-time linear mixed models in SW-CRTs with continuous recruitment under different data-generating models. In Section~\ref{sec:empirical}, we use discrete-time linear mixed models to reanalyze a completed SW-CRT with continuous recruitment. In Section~\ref{sec:discussion}, we conclude with a discussion of our findings.

\section{Considerations in Linear Mixed Models for Cross-Sectional SW-CRTs}\label{sec:setup}

\subsection{Set Up and Notation}\label{sec:notation}

We consider a standard, complete SW-CRT with $I$ clusters, $J$ equally spaced periods, and $K_{ij}$ individuals per cluster-period. To focus on the main ideas, we assume a cross-sectional design, where different individuals are observed in each period within each cluster. All clusters are in the control condition when $j = 1$ and in the intervention condition when $j = J$. Let $Y_{ijk}$ denote the observed outcome for individual $k \in \{1, \ldots, K_{ij}\}$ in cluster $i \in \{1, \ldots, I\}$ and period $j \in \{1, \ldots, J\}$, and let $Z_{ij}$ be the binary intervention indicator, which equals 1 when cluster $i$ is under intervention in period $j$ and 0 otherwise. There are $Q = J-1$ intervention sequences, with $I_q$ clusters in sequence $q \in \{1, \ldots, Q\}$ and $\sumq I_q = I$. If cluster $i$ belongs to sequence $q$, where the intervention begins in period $q+1$, then its intervention indicator vector is $\bfZ_i = (Z_{i1}, \ldots, Z_{iJ})$ with $Z_{ij} = 0$ for $1 \leq j \leq q$ and $Z_{ij} = 1$ for $q < j \leq J$. For SW-CRTs with continuous recruitment, we let $t_{ijk} \in (j-1, j]$ denote the exact recruitment time of individual $k$ in cluster $i$ and period $j$. Although the period index $j$ is uniquely determined by the recruitment time, we keep this index to facilitate the discussion of recruitment patterns across cluster-periods in Section~\ref{sec:heterogeneity}. In practice, routinely collected information, such as calendar dates or days since trial initiation, can be mapped to $t_{ijk}$, with $t_{ijk}-(j-1)$ representing the proportion of period $j$ that has elapsed at enrollment. For example, if the first period spans the initial 60 days of the trial and individual $k$ in cluster $i$ is enrolled on day 30, then $t_{i1k} = 0.5$. Similar standardization procedures can be applied when recruitment-time information is available.

\subsection{Estimands}\label{sec:estimand}

We first describe a generic linear mixed model following Li et al.\cite{Li2020} and then introduce specific model variants under discrete-time and continuous-time settings. The essential ingredients for a linear mixed model in SW-CRTs are
\begin{align}
    Y_{ijk} = \text{{secular trend}} + \text{{intervention effect}} + \text{{heterogeneity}} + \text{{residual error}}, \label{model:general}
\end{align}
where $Y_{ijk}$ is assumed to be continuous. The residual errors are independent and identically distributed as $\epsilon_{ijk} \sim \calN(0, \sigma_\epsilon^2)$. We discuss different specifications of the secular trend and heterogeneity terms under discrete sampling and continuous recruitment in Section~\ref{sec:continuous_recruitment}. The intervention effect term captures the change in the mean outcome for each treatment sequence in each period and includes the parameter of interest. For example, one may target a constant intervention effect $\delta$. Alternatively, when the intervention effect varies by exposure time, we represent it by an effect curve $\delta(s)$, where $s = j-q \in \{1, \ldots, J-1\}$ indicates the duration of exposure or exposure time. In this setting, the estimand of interest is the exposure time-averaged intervention effect:
\begin{align*}
    \Delta = \frac{1}{J-1}\sums\delta(s),
\end{align*}
where $\delta(s)$ is the point intervention effect at exposure time $s$. In the special case where the intervention effect does not change over time, $\delta(s) = \delta$ for $s \in \{1, \ldots, J-1\}$, and $\Delta = \delta$. Of note, $\Delta$ averages over all $J-1$ exposure times, following recent conventions,\cite{Kenny2022,Maleyeff2022,Wang2026} although effects averaged over a subset of exposure times may be of interest depending on the research question. We note that the intervention effect may also depend on calendar time;\cite{Wang2024,Lee2025} however, we focus on constant and exposure-time-dependent intervention effects in this article.

\subsection{Envisioning the Data-Generating Process Under Continuous Recruitment}\label{sec:continuous_recruitment}

We consider linear mixed model formulations based on \eqref{model:general} to characterize continuous recruitment. These formulations serve as the data-generating models for evaluating the performance of conventional discrete-time linear mixed models in our simulation study.

\subsubsection{Period Effect.}\label{sec:period_effect}

Because the intervention is confounded with time, accounting for the background secular trend is necessary for unbiased estimation of the intervention effect.\cite{Hussey2007,Hemming2017} Under discrete sampling, the Hussey and Hughes\cite{Hussey2007} model specifies the period effect term as
\begin{align*}
    \mu + \beta_2 \bbone(j = 2) + \ldots + \beta_J \bbone(j = J),
\end{align*}
where $\mu$ is the mean outcome under control in the first period and $\bfbeta = (\beta_2, \ldots, \beta_J)$ is the vector of secular trend parameters, with $\beta_1 = 0$ for identifiability. This saturated specification provides a nonparametric representation of the period effect and is suitable for trials with a limited number of discrete periods.\cite{Li2020} In contrast, under continuous recruitment, the period effect term can be written as\cite{Hooper2024}
\begin{align*}
    \mu + T(t_{ijk}),
\end{align*}
where $T(\cdot)$ describes the underlying continuous time trend in the expected outcome as a function of recruitment time $t_{ijk}$. Here, we assume that the same period effect function applies to all clusters.

\subsubsection{Heterogeneity.}\label{sec:heterogeneity}

In SW-CRTs, individual-level outcomes within the same cluster tend to be positively correlated.\cite{Taljaard2020} This within-cluster dependence is often characterized by the intracluster correlation coefficient (ICC)\cite{Murray2007} and must be accounted for during analysis to obtain valid statistical inference.\cite{Turner2017} Within the linear mixed model framework, different random-effects specifications induce different marginal correlation structures for the outcomes within each cluster. Three commonly used discrete-time structures are the exchangeable (EXCH), nested exchangeable (NE), and discrete-time exponential decay (DTD) structures; see Section~3.4 in Li et al.\cite{Li2020} for a review. Under the EXCH, NE, and DTD structures, the heterogeneity term in \eqref{model:general} is given by $\alpha_i$, $\alpha_i + \gamma_{ij}$, and $\gamma_{ij}$, respectively. Specifically, the EXCH structure assumes $\alpha_i \sim \calN(0, \tau_\alpha^2)$, implying a common ICC within and across periods.\cite{Hughes2015} The NE structure introduces independent cluster-by-period random effects $\gamma_{ij} \sim \calN(0, \tau_\gamma^2)$, which are also independent of the cluster-level random effect $\alpha_i$. This specification distinguishes a within-period ICC (the correlation between outcomes from two individuals in the same cluster-period) from a constant between-period ICC (the correlation between outcomes from two individuals in the same cluster but different periods).\cite{Girling2016,Hooper2016} The DTD structure assumes $(\gamma_{i1}, \ldots, \gamma_{iJ})' \sim \calN(0, \tau_\gamma^2 \bfM(r_0, r))$, where
\begin{align*}
    \bfM(r_0, r) = \left(\begin{array}{ccccc}
         1 & r_0r & r_0r^2 & \cdots & r_0r^{J-1} \\
         r_0r & 1 & r_0r & \cdots & r_0r^{J-2} \\
         \vdots & \vdots & \vdots & \ddots & \vdots \\
         r_0r^{J-1} & r_0r^{J-2} & r_0r^{J-3} & \cdots & 1
    \end{array}\right),
\end{align*}
allowing the between-period ICC to decay exponentially as the number of periods separating the measurements increases.\cite{Kasza2019a,Kasza2019b} As a side note, the choices $\bfM(1, 1)$ and $\bfM(r_0, 1)$ correspond to the EXCH and NE correlation structures, respectively. Following Kasza et al.,\cite{Kasza2019a} we use the first-order autoregressive matrix $\bfM(1, r)$ for the DTD structure in this article.

Under continuous recruitment, within-cluster correlations may depend on the exact recruitment times of individuals. We use the CTD correlation structure to characterize this dependence under different recruitment patterns. For each cluster, we index individuals within each period in chronological order. Conditional on the recruitment times, the CTD structure assumes
\begin{align*}
    \left(\gamma_{i,t_{i11}}, \ldots, \gamma_{i,t_{i1K_{i1}}}, \ldots, \gamma_{i,t_{iJ1}}, \ldots, \gamma_{i,t_{iJK_{iJ}}}\right)' \sim \calN(0, \tau_\gamma^2\widetilde{\bfM}),
\end{align*}
which is a $K_{i\cdot}$-dimensional vector of random effects ordered chronologically within cluster $i$. Here, $K_{i\cdot} = \sumj K_{ij}$ is the total number of individuals in cluster $i$ across all periods, and $\widetilde{\bfM}$ is a $K_{i\cdot} \times K_{i\cdot}$ correlation matrix. Following Grantham et al.,\cite{Grantham2019} this matrix has ones on the diagonal and off-diagonal entries $r^{|t_{ijk}-t_{ij'k'}|}$ for $(j,k) \neq (j',k')$. Thus, correlations depend on the time between measurements for individuals observed in either the same period or different periods.

\begin{figure}[t]
    \centering
    \captionsetup[subfigure]{justification=centering}
    \begin{subfigure}{0.45\textwidth}
        \centering
        \caption{Cluster Mixed Pattern}
        \includegraphics[width=\linewidth]{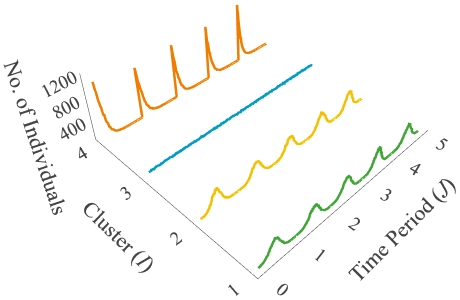}
    \end{subfigure}
    \hfill
    \begin{subfigure}{0.45\textwidth}
        \centering
        \caption{Cluster-Period Mixed Pattern}
        \includegraphics[width=\linewidth]{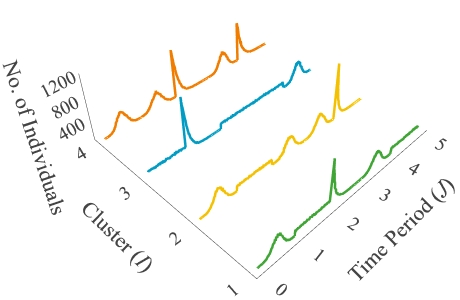}
    \end{subfigure}
    \caption{An illustration of the cluster mixed and cluster-period mixed recruitment patterns in panels (a) and (b), respectively. We consider an SW-CRT with $I = 4$ clusters, $J = 5$ periods, and $K_{ij} = 10,000$ individuals per cluster-period. The large cluster-period size is chosen to reduce sampling variability in the enrollment density surfaces. Each colored surface represents the enrollment density for one cluster over time.}
    \label{fig:patterns}
\end{figure}

We characterize recruitment patterns using the cluster-period-specific enrollment density, which describes the distribution of individual enrollment times within each period for each cluster. Motivated by the empirical recruitment patterns presented in Section~\ref{sec:empirical}, we first consider three basic patterns:
\begin{itemize}
    \item Uniform pattern: Individual enrollment times are uniformly distributed within each period, with $t_{ijk} \sim \text{Uniform}(j-1, j]$.

    \item Gaussian pattern: We generate values from $\calN(0, 1)$ and rescale them to the corresponding period using min-max normalization, producing an approximately bell-shaped enrollment pattern.

    \item Exponential pattern: We generate values from $\text{Exp}(1.5)$ and rescale them to the corresponding period using min-max normalization, producing an enrollment pattern with more individuals recruited early in each period.
\end{itemize}
To capture heterogeneity in recruitment patterns observed in real-world trials (Figure~\ref{fig:empirical_patterns}), we additionally consider two mixed patterns that combine these basic distributions:
\begin{itemize}
    \item Cluster mixed pattern: Each cluster $i$ is randomly assigned one of the three distributions (uniform, Gaussian, or exponential) with equal probability. The assigned distributional form is used to generate enrollment times in all periods within that cluster. This pattern allows recruitment to differ across clusters while maintaining the same distributional form across periods within each cluster.

    \item Cluster-period mixed pattern: Each cluster-period $(i,j)$ is independently assigned one of the three distributions (uniform, Gaussian, or exponential) with equal probability. This pattern allows recruitment to differ both across clusters and across periods within the same cluster.
\end{itemize}

Figure~\ref{fig:patterns} illustrates the differences between the cluster mixed and cluster-period mixed recruitment patterns. For illustration, we consider an SW-CRT with $I = 4$ clusters, $J = 5$ periods, and $K_{ij} = 10,000$ individuals per cluster-period, where the large cluster-period size reduces sampling variability in the enrollment densities. In Figure~\ref{fig:patterns}(a), each cluster is assigned one of the three enrollment distributions (uniform, Gaussian, or exponential), and the same distributional form is used across all periods within that cluster. This produces similar enrollment patterns across periods within each cluster while allowing the patterns to differ across clusters. In Figure~\ref{fig:patterns}(b), the enrollment distribution is selected independently for each cluster-period, allowing recruitment patterns to differ both across clusters and across periods within the same cluster. Of note, enrollment distributions are selected independently across clusters in both mixed patterns. Therefore, these patterns do not explicitly account for recruitment fluctuations shared by all clusters at the same calendar time. In practice, external calendar-time factors, such as holidays or summer months, may reduce recruitment across all clusters simultaneously, giving rise to a common temporal pattern of enrollment.

\subsubsection{Intervention Effect.}\label{sec:intervention}

To account for intervention effect heterogeneity across clusters, we also consider a random intervention effect,\cite{Hemming2018b} represented by $v_i Z_{ij}$, where $v_i \sim \calN(0, \tau_v^2)$ quantifies the departure of the cluster-specific intervention effect from the average effect $\delta$ or $\delta(s)$. In Table~\ref{table:ICC}, we summarize the ICC-related parameters under the EXCH, NE, DTD, and CTD structures, with and without a random intervention effect. This parameterization assumes that $v_i$ is independent of all other random effects, which constrains the correlation between two individuals under intervention to be at least as large as the corresponding correlation under control.\cite{Hemming2018b} We keep this independence assumption in our simulation study for simplicity and revisit it in the data application in Section~\ref{sec:empirical}. For the EXCH, NE, and DTD structures, we let $\rho_0$ and $\rho_1$ denote the within-period ICC under control and intervention, respectively. Under CTD, these parameters represent the corresponding correlations at zero time separation. Setting $\tau_v^2 = 0$ recovers the expressions without a random intervention effect, under which the ICC-related parameters do not depend on the intervention condition. We further introduce the cluster autocorrelation coefficient (CAC), defined as the ratio of the between-period ICC to the corresponding within-period ICC. Under the NE structure, the CAC is smaller than one when the variance of the random cluster-by-period interaction is positive.\cite{Hooper2016} Under the DTD structure without a random intervention effect, the CAC for adjacent periods equals $r$, the per-period correlation decay parameter.\cite{Kasza2019a}

Although recruitment times are continuous in our data-generating process, we assume a discrete-time intervention effect structure. This modeling choice is motivated by two considerations. First, allowing the intervention effect to depend on individual-level exposure time, measured from each individual's enrollment, introduces additional complexity in model specification and the definition of the target estimand. To date, there has been limited work on defining relevant and interpretable intervention effect estimands in this setting. Second, for cluster-level interventions commonly studied in SW-CRTs, provider or cluster exposure time may be more relevant than individual exposure time, especially when the intervention targets organizational practices or policies in continuous recruitment short exposure designs.\cite{Hooper2016} We therefore specify the intervention effect as a function of cluster-level exposure time, measured by the number of periods since intervention adoption. This discrete-time specification allows us to use established estimands \citep{Kenny2022,Maleyeff2022,Wang2024} as the targets in our simulation study.

\begin{table}[t]
    \caption{Summary of ICCs and CACs under the EXCH, NE, DTD, and CTD correlation structures, with and without an RI effect. The control and intervention columns refer to pairs of individuals who are both under the indicated condition. Expressions without an RI effect are obtained by setting $\tau_v^2 = 0$ and do not depend on the intervention condition.}
    \label{table:ICC}
    \renewcommand{\arraystretch}{1.3}
    \resizebox{\linewidth}{!}{
        \begin{tabular}{clccc}
            \toprule
            \multirow{2}{*}{\textbf{Correlation structure}} & \multirow{2}{*}{\textbf{Parameter}} & \multicolumn{2}{c}{\textbf{With RI effect}} & \multirow{2}{*}{\textbf{Without RI effect}} \\
            \cmidrule(lr){3-4}
            & & Control & Intervention & \\
            \midrule
            EXCH
            & ICC
            & $\dfrac{\tau_\alpha^2}{\tau_\alpha^2+\sigma_\epsilon^2}$
            & $\dfrac{\tau_\alpha^2+\tau_v^2}{\tau_\alpha^2+\tau_v^2+\sigma_\epsilon^2}$
            & $\dfrac{\tau_\alpha^2}{\tau_\alpha^2+\sigma_\epsilon^2}$ \\
            \midrule
            & Within-period ICC
            & $\dfrac{\tau_\alpha^2+\tau_\gamma^2}{\tau_\alpha^2+\tau_\gamma^2+\sigma_\epsilon^2}$
            & $\dfrac{\tau_\alpha^2+\tau_\gamma^2+\tau_v^2}{\tau_\alpha^2+\tau_\gamma^2+\tau_v^2+\sigma_\epsilon^2}$
            & $\dfrac{\tau_\alpha^2+\tau_\gamma^2}{\tau_\alpha^2+\tau_\gamma^2+\sigma_\epsilon^2}$ \\
            NE & Between-period ICC
            & $\dfrac{\tau_\alpha^2}{\tau_\alpha^2+\tau_\gamma^2+\sigma_\epsilon^2}$
            & $\dfrac{\tau_\alpha^2+\tau_v^2}{\tau_\alpha^2+\tau_\gamma^2+\tau_v^2+\sigma_\epsilon^2}$
            & $\dfrac{\tau_\alpha^2}{\tau_\alpha^2+\tau_\gamma^2+\sigma_\epsilon^2}$ \\
            & CAC
            & $\dfrac{\tau_\alpha^2}{\tau_\alpha^2+\tau_\gamma^2}$
            & $\dfrac{\tau_\alpha^2+\tau_v^2}{\tau_\alpha^2+\tau_\gamma^2+\tau_v^2}$
            & $\dfrac{\tau_\alpha^2}{\tau_\alpha^2+\tau_\gamma^2}$ \\
            \midrule
            & Within-period ICC
            & $\dfrac{\tau_\gamma^2}{\tau_\gamma^2+\sigma_\epsilon^2}$
            & $\dfrac{\tau_\gamma^2+\tau_v^2}{\tau_\gamma^2+\tau_v^2+\sigma_\epsilon^2}$
            & $\dfrac{\tau_\gamma^2}{\tau_\gamma^2+\sigma_\epsilon^2}$ \\
            DTD & Between-period ICC ($j \neq j'$)
            & $\dfrac{\tau_\gamma^2 r^{|j-j'|}}{\tau_\gamma^2+\sigma_\epsilon^2}$
            & $\dfrac{\tau_\gamma^2 r^{|j-j'|}+\tau_v^2}{\tau_\gamma^2+\tau_v^2+\sigma_\epsilon^2}$
            & $\dfrac{\tau_\gamma^2 r^{|j-j'|}}{\tau_\gamma^2+\sigma_\epsilon^2}$ \\
            & CAC ($j \neq j'$)
            & $r^{|j-j'|}$
            & $\dfrac{\tau_\gamma^2 r^{|j-j'|}+\tau_v^2}{\tau_\gamma^2+\tau_v^2}$
            & $r^{|j-j'|}$ \\
            \midrule
            & Within-period ICC
            & $\dfrac{\tau_\gamma^2}{\tau_\gamma^2+\sigma_\epsilon^2}$
            & $\dfrac{\tau_\gamma^2+\tau_v^2}{\tau_\gamma^2+\tau_v^2+\sigma_\epsilon^2}$
            & $\dfrac{\tau_\gamma^2}{\tau_\gamma^2+\sigma_\epsilon^2}$ \\
            CTD & Between-period ICC $(j,k) \neq (j',k')$
            & $\dfrac{\tau_\gamma^2 r^{|t_{ijk}-t_{ij'k'}|}}{\tau_\gamma^2+\sigma_\epsilon^2}$
            & $\dfrac{\tau_\gamma^2 r^{|t_{ijk}-t_{ij'k'}|}+\tau_v^2}{\tau_\gamma^2+\tau_v^2+\sigma_\epsilon^2}$
            & $\dfrac{\tau_\gamma^2 r^{|t_{ijk}-t_{ij'k'}|}}{\tau_\gamma^2+\sigma_\epsilon^2}$ \\
            & CAC $(j,k) \neq (j',k')$
            & $r^{|t_{ijk}-t_{ij'k'}|}$
            & $\dfrac{\tau_\gamma^2 r^{|t_{ijk}-t_{ij'k'}|}+\tau_v^2}{\tau_\gamma^2+\tau_v^2}$
            & $r^{|t_{ijk}-t_{ij'k'}|}$ \\
            \bottomrule
        \end{tabular}
    }\par\vspace{5pt}
    {\scriptsize EXCH: simple exchangeable; NE: nested exchangeable; DTD: discrete-time exponential decay; CTD: continuous-time exponential decay; RI: random intervention; ICC: intracluster correlation coefficient; CAC: cluster autocorrelation coefficient.\par}
\end{table}

\subsection{Robust Variance Estimators}\label{sec:RVE}

For discrete-time linear mixed models, we estimate the period effect parameters $(\mu, \bfbeta)$, the intervention effect parameters, and the covariance parameters from the observed data $\{(Y_{ijk}, Z_{ij}): i = 1, \ldots, I, j = 1, \ldots, J, k = 1, \ldots, K_{ij}\}$. Under the EXCH, NE, and DTD structures, the heterogeneity parameters are $\tau_\alpha^2$, $(\tau_\alpha^2, \tau_\gamma^2)$, and $(\tau_\gamma^2, r)$, respectively. We also estimate the residual variance $\sigma_\epsilon^2$ and, when a random intervention effect is included, its variance $\tau_v^2$. In general, the linear mixed model can be written as
\begin{align*}
    \bfY_i = \bfD_i\bftheta + \bfE_i\bfu_i + \bfepsilon_i,
\end{align*}
where $\bfY_i = (Y_{i11}, \ldots, Y_{iJK_{iJ}})'$ is the $K_{i\cdot} \times 1$ outcome vector for cluster $i$, and $\bfD_i$ and $\bfE_i$ are the design matrices for the fixed and random effects, respectively. Under a constant intervention effect, the fixed effect parameter vector is $\bftheta = (\mu, \bfbeta, \delta)'$. Under an exposure-time-dependent intervention effect, $\bftheta = \{\mu, \bfbeta, \delta(1), \ldots, \delta(J-1)\}'$, and the average intervention effect is estimated by $\hat\Delta = (J-1)^{-1}\sums\hat\delta(s)$. The random effect vector satisfies $\bfu_i \sim \calN(\mathbf{0}, \bfG_i)$, where $\bfG_i$ is determined by the random-effects specification in Section~\ref{sec:heterogeneity}. The residual error vector is $\bfepsilon_i = (\epsilon_{i11}, \ldots, \epsilon_{iJK_{iJ}})'$, with $\bfepsilon_i \sim \calN(\mathbf{0}, \sigma_\epsilon^2\bfI_{K_{i\cdot}})$ independently of $\bfu_i$. The working marginal covariance matrix is therefore $\bfW_i = \bfE_i\bfG_i\bfE_i' + \sigma_\epsilon^2\bfI_{K_{i\cdot}}$, where $\bfI_{K_{i\cdot}}$ is the $K_{i\cdot} \times K_{i\cdot}$ identity matrix. Let $\widehat{\bfW}_i$ denote the fitted covariance matrix and $\hat\bftheta$ the resulting fixed effect estimator. The model-based variance estimator is
\begin{align*}
    \bbV_{\text{Naive}}(\hat\bftheta) = \left(\sumi\bfD_i'\widehat{\bfW}_i^{-1}\bfD_i\right)^{-1}.
\end{align*}
Although this variance estimator is often the default in existing software, it can lead to invalid inference when the random-effects structure is misspecified.\cite{Kasza2019a,Kasza2019b} An alternative is the robust variance estimator (RVE) of Liang and Zeger,\cite{Liang1986} denoted by $\bbV_{\text{RVE}}(\hat\bftheta)$. However, the uncorrected RVE can underestimate variability and lead to inflated type I error rates when the number of clusters is small. To address this issue, Ouyang et al.\cite{Ouyang2024} recommended the Mancl and DeRouen (MD)\cite{Mancl2001} correction combined with a $t$ reference distribution with $I-2$ degrees of freedom (DoF) for SW-CRTs with fewer than 32 clusters. Specifically, the MD-corrected RVE is
\begin{align*}
    \bbV_{\text{RVE}}^{\text{MD}}(\hat\bftheta)= \left(\sumi \bfD_i'\widehat{\bfW}_i^{-1}\bfD_i\right)^{-1}
    \left(\sumi \bfD_i'\widehat{\bfW}_i^{-1}\bfA_i\bfr_i\bfr_i'\bfA_i'\widehat{\bfW}_i^{-1}\bfD_i\right)\left(\sumi \bfD_i'\widehat{\bfW}_i^{-1}\bfD_i\right)^{-1},
\end{align*}
where $\bfr_i = \bfY_i-\bfD_i\hat\bftheta$ is the residual vector for cluster $i$, $\bfA_i = (\bfI_{K_{i\cdot}}-\bfH_i)^{-1}$ is the adjustment matrix, and $\bfH_i = \bfD_i(\sumi\bfD_i'\widehat{\bfW}_i^{-1}\bfD_i)^{-1}\bfD_i'\widehat{\bfW}_i^{-1}$ is the cluster-leverage matrix. The uncorrected RVE is recovered by setting $\bfA_i = \bfI_{K_{i\cdot}}$. For exposure-time-dependent effects, the variance estimator for $\hat\Delta$ is obtained from the corresponding linear contrast of $\hat\bftheta$.

Bell and McCaffrey\cite{Bell2002} showed that this correction closely approximates the leave-one-cluster-out jackknife variance estimator under an unweighted linear model. Through simulation studies with a constant intervention effect, discrete sampling, and potentially misspecified random-effects structures, Ouyang et al.\cite{Ouyang2024} found that the MD correction combined with $I-2$ DoF achieved approximately nominal 95\% confidence interval coverage with as few as eight clusters, at the expense of a slight overestimation of standard errors. Section~3.6 in Ouyang et al.\cite{Ouyang2024} provides a detailed comparison of small-sample corrections for RVEs in SW-CRTs under discrete sampling. Implementation details for fitting each working model and computing the corresponding variance estimators are provided in Section~\ref{sce:performance}.

\section{Simulation Study}\label{sec:simulation}

In this section, we conduct a simulation study to evaluate the performance of discrete-time linear mixed model estimators in SW-CRTs with continuous recruitment. We use the framework in Section~\ref{sec:continuous_recruitment} to specify the data-generating models under continuous-time recruitment. Because discrete-time analysis is common practice in cross-sectional SW-CRTs,\cite{Nevins2024} our primary objective is to assess the reliability of discrete-time linear mixed models under the model misspecification associated with continuous-time individual recruitment. We also fit models with a CTD working correlation structure as a benchmark when computationally feasible. Table~\ref{tab:overview} summarizes the simulation scenarios for ease of reference.

\subsection{Trial Design}\label{sec:design}

We consider complete, cross-sectional SW-CRTs with $I$ clusters, $J$ equally spaced periods, $Q = J-1$ intervention sequences, and $K$ individuals recruited continuously within each cluster-period, following the setting in Section~\ref{sec:notation}. We assume a balanced design in which $I$ is divisible by $Q$, with $I_q = I/Q$ clusters assigned to sequence $q$ for $q \in \{1, \ldots, Q\}$. Clusters assigned to sequence $q$ begin receiving the intervention in period $j = q+1$. We initially assume equal cluster-period sizes and consider unequal sizes in Section~\ref{sec:sim_unequal}.

\subsection{Data-Generating Process Under Continuous Recruitment}\label{sec:data}

Within the linear mixed model framework, we characterize continuous recruitment through two complementary aspects. First, when individuals enter the study continuously, it is natural to replace the discrete period effect with a continuous function of time. Second, the correlation structure can account for the exact timing of individual outcome measurements. Specifically, the DTD structure depends on discrete period indices, whereas the CTD structure depends on pairwise differences in measurement times within clusters. For simplicity, we assume that outcomes are assessed at the enrollment encounter, so that recruitment and measurement times coincide. Of note, the same CTD correlations are obtained when all individuals have a common follow-up duration. If an individual is recruited at time $t_{ijk}$ and measured at $r_{ijk} = t_{ijk} + \Delta t$, where $\Delta t$ is the common follow-up duration, then $|r_{ijk}-r_{ij'k'}| = |t_{ijk}-t_{ij'k'}|$ for any two individuals within the same cluster. The CTD structure is therefore unchanged by this common shift from recruitment to measurement times. Motivated by the recruitment patterns observed in the Australian disinvestment trial (Figure~\ref{fig:empirical_patterns}), we consider three patterns ranging from idealized to more realistic settings: (i) uniform recruitment, where individual recruitment times are uniformly distributed within each cluster-period; (ii) cluster mixed recruitment, where the recruitment distribution varies across clusters but remains the same across periods within each cluster; and (iii) cluster-period mixed recruitment, where the recruitment distribution varies both across clusters and across periods within each cluster. All three recruitment patterns are formally defined in Section~\ref{sec:heterogeneity}.

\begin{table}[!h]
\caption{(Top) Specifications of true and working models in the simulation study. For the period effect, ``$\newcheckmark$'' indicates continuous specification and ``$\newcrossmark$'' indicates discrete specification. For the intervention effect, under the ``Time Varying'' column, ``$\newcheckmark$'' indicates an exposure-time-dependent effect and ``$\newcrossmark$'' indicates a constant effect; under the ``RI'' column, ``$\newcheckmark$'' indicates the presence of an RI effect and ``$\newcrossmark$'' indicates its absence. For the correlation structure, ``$\newcheckmark$'' indicates the presence of a specific structure and ``$\newcrossmark$'' indicates its absence. All true models assume a CTD correlation structure. Working models include three discrete-time structures (EXCH, NE, DTD) as well as the CTD structure; CTD is applied as a working model only in Simulation Study VIII under Scenario 5 to directly compare against the three discrete-time structures. (Bottom) Mapping of simulation studies to scenarios. Each cell indicates which scenarios are used in the corresponding simulation study.} \label{tab:overview}
\resizebox{\linewidth}{!}{
    \begin{tabular}{cccccccccccc}
        \toprule
        \multirow{4}{*}{\textbf{Scenario}} &
        \multicolumn{4}{c}{\textbf{True Model}} &
        \multicolumn{7}{c}{\textbf{Working Model}} \\ \cmidrule(lr){2-5} \cmidrule(lr){6-12}
        & Period Effect
        & \multicolumn{2}{c}{Intervention Effect} 
        & Correlation Structure
        & Period Effect
        & \multicolumn{2}{c}{Intervention Effect} 
        & \multicolumn{4}{c}{Correlation Structure} \\ \cmidrule(lr){2-2} \cmidrule(lr){3-4} \cmidrule(lr){5-5} \cmidrule(lr){6-6} \cmidrule(lr){7-8} \cmidrule(lr){9-12}
        & Continuous & Time Varying & RI & CTD & Continuous & Time Varying & RI & EXCH & NE & DTD & CTD \\ \midrule
        1 & \newcrossmark & \newcrossmark & \newcheckmark & \newcheckmark & \newcrossmark & \newcrossmark & \newcrossmark & \newcheckmark & \newcheckmark & \newcheckmark & \newcrossmark \\
        2 & \newcheckmark & \newcrossmark & \newcheckmark & \newcheckmark & \newcrossmark & \newcrossmark & \newcrossmark & \newcheckmark & \newcheckmark & \newcheckmark & \newcrossmark \\
        3 & \newcheckmark & \newcrossmark & \newcrossmark & \newcheckmark & \newcrossmark & \newcrossmark & \newcrossmark & \newcheckmark & \newcheckmark & \newcheckmark & \newcrossmark \\
        4 & \newcheckmark & \newcheckmark & \newcheckmark & \newcheckmark & \newcrossmark & \newcheckmark & \newcrossmark & \newcheckmark & \newcheckmark & \newcheckmark & \newcrossmark \\
        5 & \newcheckmark & \newcrossmark & \newcheckmark & \newcheckmark & Both\textsuperscript{\textdagger} & \newcrossmark & \newcrossmark & \newcheckmark & \newcheckmark & \newcheckmark & \newcheckmark \\
        \midrule[\heavyrulewidth]
        \multicolumn{12}{c}{\textbf{Simulation Study Mapping and Goals}} \\
        \midrule
        \multicolumn{2}{l}{Simulation Study I} & \multicolumn{5}{c}{Included Scenario 1 \& Scenario 2} & \multicolumn{5}{l}{Impact of period effect specification} \\
        \multicolumn{2}{l}{Simulation Study II} & \multicolumn{5}{c}{Included Scenario 2} & \multicolumn{5}{l}{Impact of recruitment patterns} \\
        \multicolumn{2}{l}{Simulation Study III} & \multicolumn{5}{c}{Included Scenario 2} & \multicolumn{5}{l}{Impact of ICC-related parameters} \\
        \multicolumn{2}{l}{Simulation Study IV} & \multicolumn{5}{c}{Included Scenario 2 \& Scenario 3} & \multicolumn{5}{l}{Impact of the random intervention effect} \\
        \multicolumn{2}{l}{Simulation Study V} & \multicolumn{5}{c}{Included Scenario 2} & \multicolumn{5}{l}{Impact of trial design parameters} \\
        \multicolumn{2}{l}{Simulation Study VI} & \multicolumn{5}{c}{Included Scenario 4} & \multicolumn{5}{l}{Impact of exposure-time-dependent treatment effect} \\
        \multicolumn{2}{l}{Simulation Study VII} & \multicolumn{5}{c}{Included Scenario 2 \& Scenario 4} & \multicolumn{5}{l}{Impact of intervention-dependent recruitment} \\
        \multicolumn{2}{l}{Simulation Study VIII} & \multicolumn{5}{c}{Included Scenario 5} & \multicolumn{5}{l}{Comparison with CTD working models} \\
        \bottomrule
    \end{tabular}
    }\par\vspace{5pt}
    {\scriptsize EXCH: simple exchangeable; NE: nested exchangeable; DTD: discrete-time exponential decay; CTD: continuous-time decay; RI: random intervention; \textsuperscript{\textdagger} Under Scenario~5, the EXCH, NE, and DTD working models use discrete period indicators. The CTD working models use discrete period indicators, the continuous oracle basis, or a continuous natural cubic spline.\par}
\end{table}

Table~\ref{tab:overview} summarizes the five scenarios considered in our simulation study. Scenario~1 assumes a discrete period effect and a constant intervention effect (i.e., the Hussey and Hughes\cite{Hussey2007} model) and adds a random intervention effect and a CTD correlation structure:
\begin{align*}
    Y_{ijk} = \mu + \beta_j + (\delta + v_i) Z_{ij} + \gamma_{i, t_{ijk}} + \epsilon_{ijk}.
\end{align*}
Scenario~2 replaces the discrete period effect $\beta_j$ with a continuous period effect $T(t_{ijk})$, with the same intervention effect and correlation specifications as Scenario~1. Scenario~3 follows Scenario~2 but excludes the random intervention effect, replacing $(\delta + v_i)Z_{ij}$ with $\delta Z_{ij}$. Scenario~4 extends Scenario~2 by replacing $\delta$ with $\delta(s)$ to allow the intervention effect to vary with exposure time. Scenario~5 uses the same data-generating process as Scenario~2 but additionally includes the CTD working model as a benchmark. Across all five scenarios, the true correlation structure is CTD, and the discrete-time working models use the EXCH, NE, or DTD structure described in Section~\ref{sec:heterogeneity}. The bottom panel of Table~\ref{tab:overview} identifies the scenarios used in each simulation study. The full specification of the data-generating process under each scenario is provided in Table~\ref{table:dgp}.

\subsection{Simulation Parameters}\label{sce:parameters}

The simulation parameters are motivated by published studies and summarized in Web Table~1. Specifically, Nevins et al.\cite{Nevins2023,Nevins2024} reported a median of 11 randomized clusters (Q1--Q3: 8--18), 5 intervention sequences (Q1--Q3: 4--7), and 2,724 participants (Q1--Q3: 643--14,734), where Q1 and Q3 denote the first and third quartiles, respectively. Additionally, Korevaar et al.\cite{Korevaar2021} examined 44 continuous outcomes from the CLustered Outcome Dataset bank and reported ICCs typically ranging from 0.01 to 0.1, with a median CAC of 0.73 (Q1--Q3: 0.19--0.91).

Based on these findings, we consider SW-CRTs with $I \in \{16, 32\}$ clusters, $J \in \{5, 9\}$ periods, $Q = J-1 \in \{4, 8\}$ intervention sequences, and $K_{ij} = 50$ individuals in each cluster-period. Under the CTD data-generating model, we fix $\sigma_\epsilon^2 = 1$ and consider settings with and without a random intervention effect. When a random intervention effect is included, we specify $\rho_0 \in \{0.01, 0.05\}$, $\rho_1 = 0.1$, and a CAC under control of 0.5 or 0.8 at a time separation of one period. Here, $\rho_0$ and $\rho_1$ denote the ICCs at zero time separation under control and intervention, respectively. In the absence of a random intervention effect, we specify $\rho_0 = \rho_1 \in \{0.01, 0.05\}$ and a CAC of 0.5 or 0.8 at the same time separation under both conditions. The corresponding variance components are obtained from the expressions in Table~\ref{table:ICC}.

To evaluate the type I error rate of the Wald test, we set $\delta = 0$ under a constant intervention effect and $\Delta = 0$ under an exposure-time-dependent intervention effect. Of note, $\Delta = 0$ requires the point intervention effects to average to zero but does not require $\delta(s) = 0$ at every exposure time. For each simulation setting, we generate 2,000 datasets, corresponding to a Monte Carlo standard error of approximately 0.5 percentage points at a coverage probability of 95\%.\cite{Morris2019}

\begin{table}[t]
\caption{Specification of the true data-generating process under each scenario in the simulation study. Under the random intervention (RI) column, ``$\newcheckmark$'' indicates the presence of a random intervention effect and ``$\newcrossmark$'' indicates its absence.} \label{table:dgp}
\resizebox{\linewidth}{!}{
    \begin{tabular}{clccc}
        \toprule
        \textbf{Scenario} & \textbf{Data-Generating Model} & \textbf{Period Effect} & \textbf{Intervention Effect} & \textbf{RI} \\
        \midrule
        1 & $Y_{ijk} = \mu + \beta_j + (\delta + v_i)Z_{ij} + \gamma_{i, t_{ijk}} + \epsilon_{ijk}$
          & $\beta_j = 0.5j^{2}/J$ (discrete)
          & $\delta$ (constant)
          & \newcheckmark \\
        \addlinespace
        2 & $Y_{ijk} = \mu + T(t_{ijk}) + (\delta + v_i)Z_{ij} + \gamma_{i, t_{ijk}} + \epsilon_{ijk}$
          & $T(t_{ijk}) = \{0.5t_{ijk}^{2} + \sin(6\pi t_{ijk})\}/J$ (continuous)
          & $\delta$ (constant)
          & \newcheckmark \\
        \addlinespace
        3 & $Y_{ijk} = \mu + T(t_{ijk}) + \delta Z_{ij} + \gamma_{i, t_{ijk}} + \epsilon_{ijk}$
          & $T(t_{ijk}) = \{0.5t_{ijk}^{2} + \sin(6\pi t_{ijk})\}/J$ (continuous)
          & $\delta$ (constant)
          & \newcrossmark \\
        \addlinespace
        4 & $Y_{ijk} = \mu + T(t_{ijk}) + \{\delta(s) + v_i\}Z_{ij} + \gamma_{i, t_{ijk}} + \epsilon_{ijk}$
          & $T(t_{ijk}) = \{0.5t_{ijk}^{2} + \sin(6\pi t_{ijk})\}/J$ (continuous)
          & $\delta(s)$ (exposure-time-dependent)
          & \newcheckmark \\
        \addlinespace
        5 & $Y_{ijk} = \mu + T(t_{ijk}) + (\delta + v_i)Z_{ij} + \gamma_{i, t_{ijk}} + \epsilon_{ijk}$
          & $T(t_{ijk}) = \{0.5t_{ijk}^{2} + \sin(6\pi t_{ijk})\}/J$ (continuous)
          & $\delta$ (constant)
          & \newcheckmark \\
        \bottomrule
    \end{tabular}
}
\end{table}

\subsection{Data Analysis and Performance Measures}\label{sce:performance}

All simulated datasets are analyzed at the individual level using linear mixed models in R. The working models correctly specify whether the intervention effect is constant or depends on exposure time.We specify a discrete period effect under the EXCH, NE, and DTD working structures and consider discrete and continuous period-effect specifications under CTD. Although the data-generating process includes a random intervention effect in all scenarios except Scenario~3, the working models include only fixed intervention effect terms. We obtain $\hat\bftheta$ and $\bbV_{\text{Naive}}(\hat\bftheta)$ using the \texttt{lmerTest} package under EXCH and NE, the \texttt{glmmTMB} package under DTD, and the \texttt{glmmrBase} package under CTD. The EXCH, NE, and DTD working models are fitted using restricted maximum likelihood, whereas the CTD working model is fitted using Markov chain Monte Carlo maximum likelihood.\cite{Samuel2024} For CTD, we restrict the evaluation to the model-based variance estimator.

For the three discrete-time working structures, we evaluate both the uncorrected RVE and the RVE with the MD correction. Under EXCH and NE, we compute these estimators using the \texttt{clubSandwich} package, with ``CR0'' for the uncorrected RVE and ``CR3'' for the MD-corrected RVE. The CR3 correction closely approximates the leave-one-cluster-out jackknife variance estimator.\cite{Bell2002} Under DTD, we compute the MD-corrected RVE using the R function developed by Wang et al.\cite{Wang2024}

A complete list of performance measures is provided in Web Table~2. To account for the small number of clusters, we construct 95\% confidence intervals and conduct two-sided Wald tests using a $t$ reference distribution with $I-2$ degrees of freedom. This choice of degrees of freedom was recommended by Ford and Westgate\cite{Ford2020} for GEE estimators and has shown good performance in simulation studies of SW-CRTs.\cite{Li2019,Li2021,DavisPlourde2021,Ouyang2024} However, combining this reference distribution with the MD correction can lead to conservative inference in some settings.

\subsection{Simulation Results}\label{sec:sim_result}

In each simulation study, we vary one feature of the true data-generating process while holding the remaining features fixed, allowing us to address a specific question about the performance of the discrete-time working models (Table~\ref{tab:overview} and Web Table~3). Across all simulation studies, we report point estimates and standard errors to three decimal places, empirical coverage probabilities (\%) to one decimal place, and convergence rates (\%) as integers.

\subsubsection{Simulation Study I (Scenarios 1--2): Impact of the Period Effect Specification in the Data-Generating Process.}\label{sec:sim_period_effect}

In Simulation Study~I, we examine whether replacing a discrete period effect with a continuous period effect in the data-generating process affects the performance of discrete-time linear mixed models. Specifically, we compare Scenarios~1 and~2 (Table~\ref{tab:overview}), which differ only in the period effect: Scenario~1 specifies $\beta_j = 0.5 j^2/J$, whereas Scenario~2 specifies $T(t_{ijk}) = \{0.5 t_{ijk}^{2} + \sin(6\pi t_{ijk})\}/J$. Both specifications include a quadratic trend, with additional sinusoidal variation in Scenario~2 to mimic seasonal fluctuations. Both scenarios assume a constant average intervention effect and include a random intervention effect. We consider a standard SW-CRT with $I = 32$, $J = 5$, and $K_{ij} = 50$. Data are generated under a CTD correlation structure with cluster-period mixed recruitment. We specify $\rho_0 = 0.01$, $\rho_1 = 0.1$, $\sigma_\epsilon^{2} = 1$, and a CAC under control of 0.5 at a time separation of one period. We set $\delta = 0$ to evaluate type I error rates and analyze each simulated dataset using discrete-time linear mixed models with EXCH, NE, and DTD working correlation structures.

\begin{table}[!htbp]
    \caption{Simulation results based on 2,000 datasets generated under Scenarios 1 and 2 for Simulation Studies I and II. The design assumes a standard SW-CRT with $I = 32$, $J = 5$, $K_{ij} = 50$, $\delta = 0$, $\rho_0 = 0.01$, $\rho_1 = 0.1$, $\sigma_\epsilon^{2} = 1$, and CAC $= 0.5$. The discrete period effect is set to $\beta_j = 0.5 j^2/J$, and the continuous period effect is set to $T(t_{ijk}) = \{0.5 t_{ijk}^{2} + \sin(6\pi t_{ijk})\}/J$. $\sd(\cdot)$ denotes the empirical (Monte Carlo) standard deviation. $\se_{\text{Naive}}(\cdot)$, $\se_{\text{RVE}}(\cdot)$, and $\se_{\text{RVE}}^{\text{MD}}(\cdot)$ denote the average model-based, robust, and MD-corrected robust standard errors, respectively. $\bbC_{\text{Naive}}(\cdot)$, $\bbC_{\text{RVE}}(\cdot)$, and $\bbC_{\text{RVE}}^{\text{MD}}(\cdot)$ denote the corresponding empirical coverage of the 95\% confidence intervals.}\label{tab:I_and_II}
    \resizebox{\linewidth}{!}{
    \begin{tabular}{cclccccccccc}
        \toprule
        Period Effect & Pattern & Model & Bias & sd$(\hat\delta)$ & $\se_{\text{Naive}}(\hat\delta)$ & $\bbC_{\text{Naive}}(\hat\delta)$ & $\se_{\text{RVE}}(\hat\delta)$ & $\bbC_{\text{RVE}}(\hat\delta)$ & $\se_{\text{RVE}}^{\text{MD}}(\hat\delta)$ & $\bbC_{\text{RVE}}^{\text{MD}}(\hat\delta)$ & Convergence \\ 
        \midrule
        \multirow{3}{*}{Discrete} & \multirow{3}{*}{CP} & EXCH & 0.001 & 0.073 & 0.040 & 73.6 & 0.068 & 93.2 & 0.073 & 95.0 & 100 \\
        & & NE & 0.001 & 0.073 & 0.059 & 89.5 & 0.068 & 93.2 & 0.073 & 95.2 & 99 \\
        & & DTD & 0.001 & 0.073 & 0.057 & 88.5 & 0.068 & 94.0 & 0.072 & 95.0 & 100 \\
        \midrule
        \multirow{9}{*}{Continuous} & \multirow{3}{*}{CP} & EXCH & 0.001 & 0.074 & 0.040 & 73.9 & 0.070 & 93.7 & 0.075 & 95.5 & 100 \\
        & & NE & 0.001 & 0.074 & 0.061 & 89.8 & 0.070 & 93.6 & 0.074 & 95.2 & 99 \\
        & & DTD & 0.001 & 0.074 & 0.059 & 89.2 & 0.069 & 93.8 & 0.074 & 95.3 & 100 \\
        \cmidrule{2-12}
         & \multirow{3}{*}{C} & EXCH & 0.001 & 0.073 & 0.041 & 73.8 & 0.069 & 94.7 & 0.073 & 95.8 & 100 \\
        & & NE & 0.001 & 0.073 & 0.061 & 91.2 & 0.069 & 94.2 & 0.073 & 95.5 & 99 \\ 
        & & DTD & 0.001 & 0.072 & 0.058 & 90.5 & 0.068 & 94.1 & 0.073 & 95.7 & 100 \\ 
        \cmidrule{2-12}
         & \multirow{3}{*}{U} & EXCH & 0.000 & 0.072 & 0.040 & 74.3 & 0.068 & 94.1 & 0.073 & 95.5 & 100 \\
        & & NE & 0.000 & 0.072 & 0.059 & 90.3 & 0.068 & 93.4 & 0.073 & 95.2 & 99 \\
        & & DTD & -0.000 & 0.072 & 0.057 & 88.8 & 0.068 & 93.9 & 0.072 & 95.0 & 100 \\
        \bottomrule
    \end{tabular}
    }\par\vspace{5pt}
    {\scriptsize EXCH: simple exchangeable; NE: nested exchangeable; DTD: discrete-time exponential decay; U: uniform pattern; C: cluster mixed pattern; CP: cluster-period mixed pattern; RVE: robust variance estimator; MD: Mancl and DeRouen.\par}
\end{table}

Table~\ref{tab:I_and_II} shows that the discrete-time working models perform similarly under the two period-effect specifications. Across the EXCH, NE, and DTD working correlation structures, $\hat\delta$ has negligible bias under both the discrete period effect in Scenario~1 and the continuous period effect in Scenario~2. These findings are consistent with Wang et al.,\cite{Wang2024} which showed that the intervention effect estimator remains consistent under misspecification of the period effect, random-effects structure, or residual error distribution when the intervention effect structure is correctly specified. Although their theory was developed under a discrete-time framework, our results provide empirical support for similar robustness under continuous recruitment. We therefore focus subsequent comparisons on variance estimation. When confidence intervals are constructed using the model-based variance estimator, coverage is lowest under EXCH and improves to approximately 89--90\% under NE and DTD, but remains below the nominal 95\% level for all three working structures. Using the uncorrected RVE increases coverage to 93--94\% across all three working structures, with slight undercoverage. Among the variance estimators considered, only the MD-corrected RVE achieves approximately nominal coverage under all three working structures in both scenarios, supporting the use of this correction for inference in these settings. Because the discrete and continuous period effects yield similar results, we use the continuous specification in subsequent data-generating processes. Convergence rates exceed 98\% in most settings, and we report rates below this threshold for the corresponding settings.

\subsubsection{Simulation Study II (Scenario 2): Impact of Recruitment Patterns in the Data-Generating Process.}\label{sec:sim_pattern}

In Simulation Study~II, we compare uniform, cluster mixed, and cluster-period mixed recruitment under Scenario~2, with all other settings as in Simulation Study~I. Each simulated dataset is analyzed using EXCH, NE, and DTD working models. Table~\ref{tab:I_and_II} shows that recruitment patterns have little influence on point estimation or variance estimation. Although the empirical standard deviation of $\hat\delta$ is slightly larger under cluster-period mixed recruitment, the conclusions regarding bias and confidence interval coverage remain unchanged. We therefore use this recruitment pattern in subsequent simulation studies (Studies~III--VI).

\subsubsection{Simulation Study III (Scenario 2): Impact of ICC-Related Parameters in the Data-Generating Process.}\label{sec:sim_ICC}

In Simulation Study~III, we examine how the CAC and ICC parameter $\rho_0$ in the data-generating process affect the performance of discrete-time working models. We use the cluster-period mixed recruitment setting from Simulation Study~II and vary the CAC under control over $\{0.5, 0.8\}$ and $\rho_0$ over $\{0.01, 0.05\}$, with all other parameters fixed. Table~\ref{tab:CAC_and_ICC} shows that the CAC has little influence on point estimation or variance estimation. In contrast, increasing $\rho_0$ from 0.01 to 0.05 improves the coverage of confidence intervals based on the model-based variance estimators under NE and DTD from approximately 88\%--89\% to 92--93\% for a fixed CAC. Across all parameter combinations, confidence intervals based on the uncorrected RVE achieve approximately 93--94\% coverage, while those based on the MD-corrected RVE achieve nominal coverage.

\begin{table}[t]
    \caption{Simulation results based on 2,000 datasets generated under Scenario 2 for Simulation Study III. The design assumes a standard SW-CRT with $I = 32$, $J = 5$, $K_{ij} = 50$, $\delta = 0$, $\rho_0 \in \{0.01, 0.05\}$, $\rho_1 = 0.1$, $\sigma_\epsilon^{2} = 1$, CAC $\in \{0.5, 0.8\}$, a continuous period effect $T(t_{ijk}) = \{0.5 t_{ijk}^{2} + \sin(6\pi t_{ijk})\}/J$, and a cluster-period mixed recruitment pattern. $\sd(\cdot)$ denotes the empirical (Monte Carlo) standard deviation. $\se_{\text{Naive}}(\cdot)$, $\se_{\text{RVE}}(\cdot)$, and $\se_{\text{RVE}}^{\text{MD}}(\cdot)$ denote the average model-based, robust, and MD-corrected robust standard errors, respectively. $\bbC_{\text{Naive}}(\cdot)$, $\bbC_{\text{RVE}}(\cdot)$, and $\bbC_{\text{RVE}}^{\text{MD}}(\cdot)$ denote the corresponding empirical coverage of the 95\% confidence intervals.}\label{tab:CAC_and_ICC}
    \resizebox{\linewidth}{!}{
    \begin{tabular}{cclccccccccc}
        \toprule
        CAC & $\rho_0$ & Model & Bias & sd$(\hat\delta)$ & $\se_{\text{Naive}}(\hat\delta)$ & $\bbC_{\text{Naive}}(\hat\delta)$ & $\se_{\text{RVE}}(\hat\delta)$ & $\bbC_{\text{RVE}}(\hat\delta)$ & $\se_{\text{RVE}}^{\text{MD}}(\hat\delta)$ & $\bbC_{\text{RVE}}^{\text{MD}}(\hat\delta)$ & Convergence\\ 
        \midrule
        \multirow{6}{*}{0.5} & \multirow{3}{*}{0.01} & EXCH & 0.001 & 0.074 & 0.040 & 73.9 & 0.070 & 93.7 & 0.075 & 95.5 & 100 \\ 
        & & NE & 0.001 & 0.074 & 0.061 & 89.8 & 0.070 & 93.6 & 0.074 & 95.2 & 99 \\ 
        & & DTD & 0.001 & 0.074 & 0.059 & 89.2 & 0.069 & 93.8 & 0.074 & 95.3 & 100 \\ 
        \cmidrule{2-12}
         & \multirow{3}{*}{0.05} & EXCH & 0.000 & 0.076 & 0.041 & 72.4 & 0.072 & 94.1 & 0.077 & 95.3 & 100 \\
        & & NE & 0.001 & 0.076 & 0.067 & 92.9 & 0.071 & 93.5 & 0.076 & 95.2 & 99 \\ 
        & & DTD & 0.001 & 0.074 & 0.065 & 92.7 & 0.069 & 93.5 & 0.074 & 95.0 & 100  \\ 
        \midrule
        \multirow{6}{*}{0.8} & \multirow{3}{*}{0.01} & EXCH & 0.000 & 0.074 & 0.041 & 73.9 & 0.068 & 93.0 & 0.073 & 94.7 & 100 \\ 
        & & NE & 0.001 & 0.074 & 0.060 & 89.0 & 0.068 & 93.0 & 0.073 & 94.7 & 99 \\ 
        & & DTD & 0.001 & 0.074 & 0.058 & 88.0 & 0.068 & 93.0 & 0.073 & 94.7 & 100 \\ 
        \cmidrule{2-12}
         & \multirow{3}{*}{0.05} & EXCH & 0.001 & 0.070 & 0.041 & 76.5 & 0.066 & 93.7 & 0.070 & 95.4 & 100 \\
        & & NE & 0.001 & 0.070 & 0.063 & 93.2 & 0.066 & 93.8 & 0.070 & 95.2 & 100 \\
        & & DTD & 0.001 & 0.069 & 0.059 & 92.0 & 0.064 & 93.8 & 0.068 & 95.4 & 100 \\
        \bottomrule
    \end{tabular}
    }\par\vspace{5pt}
    {\scriptsize EXCH: simple exchangeable; NE: nested exchangeable; DTD: discrete-time exponential decay; ICC: intracluster correlation coefficients; CAC: cluster autocorrelation coefficient; $\rho_0$: within‑period ICC under control; $\rho_1$: within‑period ICC under treatment; RVE: robust variance estimator; MD: Mancl and DeRouen.\par}
\end{table}

\subsubsection{Simulation Study IV (Scenario 2--3): Impact of the Random Intervention Effect in the Data-Generating Process.}\label{sec:sim_RI}

In Simulation Study~IV, we examine how the presence of a random intervention effect in the data-generating process affects discrete-time analysis. We use the setting from Simulation Study~III but generate data under Scenario~3, which assumes a constant intervention effect without a random intervention effect. We therefore specify $\rho_0 = \rho_1 \in \{0.01, 0.05\}$ and a CAC of 0.5 or 0.8 under both conditions. Comparing Scenario~3 (Web Table~4) with Scenario~2 (Table~\ref{tab:CAC_and_ICC}) shows that omitting the random intervention effect leads to notable changes in model-based confidence interval coverage. Under Scenario~3, confidence intervals based on the model-based variance estimators are conservative under NE and DTD, with coverage generally near or slightly above 95\%, whereas coverage under EXCH ranges from 81\% to 94\%, depending on the correlation parameters. Across all three working structures, confidence intervals based on the uncorrected RVE achieve approximately 93--94\% coverage, while those based on the MD-corrected RVE achieve nominal coverage. The convergence rate under DTD decreases to 97\% when CAC $= 0.8$ and $\rho_0 = 0.01$. These results suggest that the presence of a random intervention effect primarily affects model-based variance estimation, with the direction and magnitude depending on the working correlation structure. The MD-corrected RVE achieves nominal coverage both with and without a random intervention effect in the settings considered.

\subsubsection{Simulation Study V (Scenario 2): Impact of Trial Design Parameters.}\label{sec:sim_design}

In Simulation Study~V, we examine whether the earlier findings extend to a smaller trial with $I = 16$ and $J = 5$ (Web Table~5) and a larger trial with $I = 32$ and $J = 9$ (Web Table~6), with all other settings as in Simulation Study~III. In the smaller trial, the empirical standard deviation of $\hat\delta$ is larger than in Simulation Study~III, but the relative performance of the working models and variance estimators remains largely unchanged. In the larger trial, confidence intervals based on the model-based variance estimators have lower coverage, with coverage lowest under EXCH, intermediate under NE, and highest under DTD. This finding suggests that model-based inference can be more sensitive to correlation misspecification in trials with more periods. Under both trial settings, confidence intervals based on the uncorrected RVE achieve approximately 93--94\% coverage, while those based on the MD-corrected RVE achieve nominal coverage. Convergence rates exceed 98\% in all settings except NE with $J = 9$, where they range from 92\% to 95\%.

\subsubsection{Simulation Study VI (Scenario 4): Impact of an Exposure-Time-Dependent Intervention Effect in the Data-Generating Process.}\label{sec:sim_TVE}

In Simulation Study~VI, we examine the performance of discrete-time working models under an exposure-time-dependent intervention effect. We use the reference setting of Study~III with $\rho_0 = 0.01$ and CAC $= 0.5$, and generate data under Scenario~4 with $\delta(1) \approx -1.38$, $\delta(2) \approx 0.38$, $\delta(3) \approx 0.98$, $\delta(4) \approx 0.02$, and $\Delta = 0$. The working models correctly specify the exposure-time-dependent intervention effect structure. Web Table~7 summarizes the results for all four point effects $\delta(s)$ and the time-averaged effect $\Delta$. The empirical standard deviation of $\hat\delta(s)$ increases with exposure time because fewer observations available at later exposure times. Consistent with the earlier studies, confidence intervals based on the model-based variance estimators have the lowest coverage under EXCH for all estimands. Model-based coverage is higher under NE and DTD but remains below the nominal level. Across all estimands and working structures, confidence intervals based on the uncorrected RVE achieve approximately 93--95\% coverage, while those based on the MD-corrected RVE achieve nominal coverage.

\subsubsection{Simulation Study VII (Scenarios 2 and 4): Impact of Intervention-Dependent Recruitment in the Data-Generating Process.}\label{sec:sim_unequal}

In Simulation Study~VII, we examine the performance of discrete-time working models when recruitment depends on the intervention condition. This setting reflects trials where intervention implementation may increase cluster-period sizes or change recruitment patterns within periods. We first consider a constant intervention effect $\delta = 0$ under Scenario~2 with $I = 96$, $J = 5$, $\rho_0 = 0.01$, $\rho_1 = 0.1$, and CAC $= 0.5$. We consider three cluster-period size configurations: (i) balanced recruitment with $K_{ij} = 50$ under both conditions, (ii) moderately unbalanced recruitment with $K_{ij} = 25$ under control and $75$ under intervention, and (iii) severely unbalanced recruitment with $K_{ij} = 10$ under control and $90$ under intervention. For each size configuration, we compare uniform recruitment under both conditions (U + U), uniform recruitment under control and cluster-period mixed recruitment under intervention (U + CP), and cluster-period mixed recruitment under both conditions (CP + CP). Each simulated dataset is analyzed using discrete-time linear mixed models with EXCH, NE, and DTD working correlation structures.

Web Table~8 shows that U + U yields negligible bias under all three working structures, with nominal coverage for confidence intervals based on the MD-corrected RVE. Under CP + CP, bias is generally small but reaches approximately $-0.014$ to $-0.015$ in the severely unbalanced configuration. Under U + CP, non-negligible bias occurs across all three size configurations. For example, with balanced cluster-period sizes, the EXCH estimator has an empirical bias of approximately $-0.030$. Confidence intervals based on the MD-corrected RVE also fall below nominal coverage under U + CP. Across the configurations considered, the average standard error estimated using the MD-corrected RVE is close to the empirical standard deviation of the intervention effect estimator, suggesting that the undercoverage is primarily caused by bias in point estimation. This is conceptually related to the non-informative enrollment assumption used by Wang et al.,\cite{Wang2024} which requires individual-level enrollment indicators to be independent of potential outcomes, baseline covariates, and treatment assignment, conditional on cluster source population size. Intervention-dependent cluster-period sizes can violate this requirement. These findings are particularly relevant to trials susceptible to identification and recruitment bias,\cite{Easter2021} where knowledge of the intervention condition may influence participant identification and enrollment.

Web Table~9 extends the analysis by varying the number of clusters, $I \in \{16, 32, 96\}$, and allowing cluster-period sizes to depend on exposure time under $\delta = 0$. We consider two size configurations: (i) S1 with $K_{ij} = 25$ under control and $K_{ij} = 50, 75, 100, 125$ at exposure times 1--4, respectively, and (ii) S2 with $K_{ij} = 10$ under control and $K_{ij} = 100, 110, 130, 160$ at exposure times 1--4, respectively. For each configuration, we consider U + U and U + CP. Under U + U, bias is negligible and confidence intervals based on the MD-corrected RVE achieve nominal coverage. Under U + CP, the empirical bias is approximately $-0.030$ across the cluster numbers and size configurations considered. We additionally examine $\delta = 2$ under $I = 96$, S2, and U + CP. The empirical bias differs little from that under $\delta = 0$, suggesting limited sensitivity to the intervention effect size in this configuration.

Web Tables~10--12 extend the analysis to exposure-time-dependent intervention effects under Scenario~4 with $I \in \{16, 32, 96\}$. Under U + U, bias is negligible for all point intervention effect estimators and the time-averaged effect estimator. Under U + CP, these estimators have non-negligible biases of similar magnitude to those observed under a constant intervention effect. The convergence rate for the NE working model is approximately 95\% when $I = 96$ (Web Table~12).

\subsubsection{Simulation Study VIII (Scenario 5): Comparison with CTD Working Models.}\label{sec:sim_CTD}

In Simulation Studies~I--VII, we evaluated the EXCH, NE, and DTD working correlation structures using data generated under a CTD correlation structure, without fitting CTD working models. Simulation Study~VIII addresses two questions: (i) how the discrete-time working models perform relative to CTD working models and (ii) how computationally demanding CTD fitting is under the implementation and trial settings considered. We consider three CTD working models that differ in their period-effect specification: discrete period indicators, an oracle basis $\{t_{ijk}^{2}, \sin(6\pi t_{ijk})\}$ matching the functional form of the true period effect, and a natural cubic spline with four degrees of freedom. We examine a smaller configuration with $I = 16$ and $K_{ij} = 20$ and a larger configuration with $I = 32$ and $K_{ij} = 50$, both under Scenario~5 (Table~\ref{tab:overview}) with $J = 5$, $\delta = 0$, $\rho_0 = 0.01$, $\rho_1 = 0.1$, $\sigma_\epsilon^2 = 1$, CAC $= 0.5$ under control, cluster-period mixed recruitment, and a random intervention effect. As in the earlier studies, the working models omit the random intervention effect. For the CTD implementation used in this study, we report model-based standard errors and the average fit time per replicate.

Web Table~13 shows that the CTD working model with discrete period indicators yields negligible bias and model-based 95\% CI coverage close to that under NE and DTD. At $I = 32$, coverage is 90.1\% for this CTD working model, compared with 75.5\% under EXCH. The oracle CTD working model also yields negligible bias. At $I = 32$, however, its empirical standard deviation is substantially larger and its model-based coverage is lower than those of the CTD working model with discrete period indicators. This increase in empirical standard deviation is not observed at $I = 16$. The natural cubic spline specification produces positive bias, including an empirical bias of $0.039$ with an average estimated standard error of $0.055$. Discrete period indicators are therefore a practical choice among the mean-model specifications considered here, yielding negligible bias without requiring knowledge of the true period-effect form.

With the implementation and settings used in Simulation Study~VIII, fitting CTD working models is computationally expensive. At $I = 32$ and $K_{ij} = 50$, a CTD fit requires approximately one hour on average, compared with less than one second for each discrete-time working model. These results demonstrate a substantial computational advantage for the discrete-time working models in the settings considered. Together with the earlier findings under intervention-independent recruitment, they support the use of discrete-time linear mixed models with the MD-corrected RVE as a practical approach to analyzing SW-CRTs with continuous recruitment.

\begin{figure}[htbp!]
    \centering
    \includegraphics[width=0.75\linewidth]{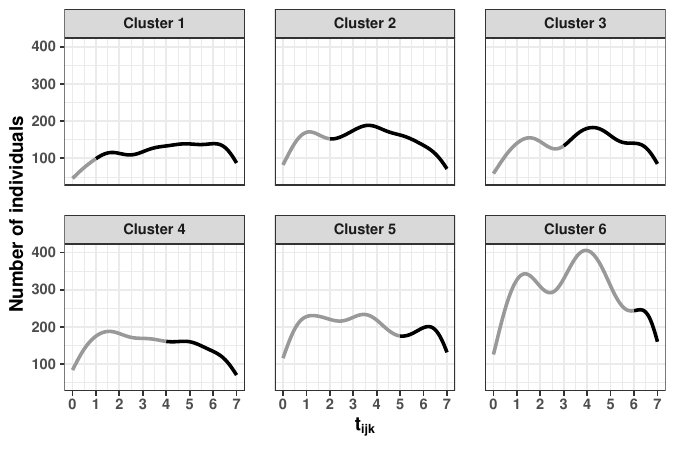}
    \caption{Empirical recruitment patterns for the Australian Disinvestment SW-CRT, based on available individual recruitment-time data from the six wards in the first hospital (batch~1). These wards contributed data during calendar periods 1--7 (Web Table~14). Recruitment-time information was unavailable for the six wards in the second hospital (batch~2). Gray lines represent control periods with the existing weekend service, and black lines represent intervention periods following service removal. The $x$-axis shows standardized recruitment time $t_{ijk}$ on a common 0--7 scale, with each unit representing one monthly period.}
    \label{fig:empirical_patterns}
\end{figure}

\section{An Illustrative Example: Reanalysis of the Australian Disinvestment Trial}\label{sec:empirical}

We analyze data from the Australian Disinvestment study, which evaluated weekend allied health services across 12 acute medical or surgical wards in two hospitals.\cite{Haines2017} This trial evaluated the impact of removing the existing weekend services, including physical therapy, occupational therapy, and social work. The analysis includes $14{,}834$ patients after excluding those exposed to both the no-service and a service condition, following the original eligibility criteria.\cite{Haines2017} The study followed a batched SW-CRT design, with each hospital forming a batch of six wards randomized to crossover sequences within that hospital.\cite{Kasza2022} Each batch was followed over seven monthly periods, corresponding to calendar periods 1--7 for the first hospital (batch~1) and 3--9 for the second hospital (batch~2). Web Table~14 groups wards by hospital and orders them by crossover time within each hospital. Although our simulations focus on standard SW-CRTs, the challenge of approximating within-cluster correlations under continuous recruitment using discrete-time working models also arises within each batch of a batched design. We therefore use this trial to illustrate the same modeling and robust variance estimation approach, including batch-specific period effects to reflect the batched design.\cite{Kasza2022}

We specifically examine length of hospital stay, measured in days, to assess patient flow following service removal. Our outcome analyses include all 12 wards from both hospitals. However, individual recruitment-time information was available only for the six wards in the first hospital (batch~1), with no recruitment-time information available for the six wards in the second hospital (batch~2). Figure~\ref{fig:empirical_patterns} therefore uses the available recruitment-time data from batch~1. The substantial variation in recruitment patterns across these wards and periods motivates the cluster-period mixed pattern described in Section~\ref{sec:heterogeneity}. Because recruitment times are unavailable for the full analysis dataset, we do not fit a CTD working model. We instead consider discrete-time linear mixed models under the EXCH, NE, and DTD working correlation structures. All models include separate period effects for each hospital, allowing each batch to have its own baseline outcome level and secular trend, with common intervention effect parameters across hospitals. 

To assess whether the intervention effect varies with exposure time, we compare the constant and exposure-time-dependent models using likelihood ratio tests. These models assume no random intervention effect and are fitted by maximum likelihood. Specifically, we test $H_0:\delta(1)=\cdots=\delta(6)$ using a chi-squared reference distribution with five degrees of freedom, obtaining $p$-values of $0.0065$, $0.0926$, and $0.1731$ under EXCH, NE, and DTD, respectively. These results suggest that the intervention effect may vary with exposure time, consistent with Haines et al.\cite{Haines2017} that staff may require an accommodation period following service withdrawal. We therefore present results under both constant and exposure-time-dependent intervention effect specifications, with $\Delta = \sum_{s=1}^{6}\delta(s)/6$ as the estimand under the latter specification. For estimation, we fit all models by restricted maximum likelihood and report model-based, uncorrected robust, and MD-corrected robust standard errors. All 95\% CIs use a $t$ distribution with $I-2=10$ degrees of freedom.

\begin{table}[!htbp]
\caption{Estimated intervention effects (in days), standard errors, and 95\% CIs under the EXCH, NE, and DTD working correlation structures for the Australian Disinvestment Trial. All models include hospital-specific period effects and assume no random intervention effect. The cross indicates a constant intervention effect, and the check indicates an exposure-time-dependent intervention effect.}\label{table:empirical}
\resizebox{\linewidth}{!}{
    \begin{tabular}{clclcc}
        \toprule
        \textbf{Time Varying} & \textbf{Working Model} & \textbf{Estimate} & \textbf{Variance Estimator} & \textbf{Estimated Standard Error} & \textbf{95\% CI} \\
        \midrule
        \multirow{9}{*}{$\newcrossmark$} & \multirow{3}{*}{EXCH} & \multirow{3}{*}{$1.65$} & Model-based & $0.25$ & $(1.10, 2.20)$ \\
         &  &  & RVE & $0.31$ & $(0.96, 2.34)$ \\
         &  &  & RVE + MD & $0.42$ & $(0.71, 2.59)$ \\
        \cmidrule{2-6}
         & \multirow{3}{*}{NE} & \multirow{3}{*}{$1.70$} & Model-based & $0.38$ & $(0.85, 2.55)$ \\
         &  &  & RVE & $0.33$ & $(0.97, 2.43)$ \\
         &  &  & RVE + MD & $0.44$ & $(0.71, 2.69)$ \\
        \cmidrule{2-6}
         & \multirow{3}{*}{DTD} & \multirow{3}{*}{$1.85$} & Model-based & $0.41$ & $(0.94, 2.76)$ \\
         &  &  & RVE & $0.29$ & $(1.19, 2.51)$ \\
         &  &  & RVE + MD & $0.40$ & $(0.97, 2.73)$ \\
        \midrule
        \multirow{9}{*}{$\newcheckmark$} & \multirow{3}{*}{EXCH} & \multirow{3}{*}{$2.56$} & Model-based & $0.39$ & $(1.69, 3.42)$ \\
         &  &  & RVE & $0.36$ & $(1.74, 3.37)$ \\
         &  &  & RVE + MD & $0.48$ & $(1.48, 3.63)$ \\
        \cmidrule{2-6}
         & \multirow{3}{*}{NE} & \multirow{3}{*}{$2.58$} & Model-based & $0.58$ & $(1.30, 3.86)$ \\
         &  &  & RVE & $0.39$ & $(1.71, 3.45)$ \\
         &  &  & RVE + MD & $0.49$ & $(1.49, 3.67)$ \\
        \cmidrule{2-6}
         & \multirow{3}{*}{DTD} & \multirow{3}{*}{$2.84$} & Model-based & $0.74$ & $(1.20, 4.48)$ \\
         &  &  & RVE & $0.54$ & $(1.63, 4.05)$ \\
         &  &  & RVE + MD & $0.76$ & $(1.14, 4.55)$ \\
        \bottomrule
    \end{tabular}
    }\par\vspace{5pt}
    {\scriptsize EXCH: simple exchangeable; NE: nested exchangeable; DTD: discrete-time exponential decay; RVE: robust variance estimator; MD: Mancl and DeRouen; CI: confidence interval.\par}
\end{table}

Estimation results are shown in Table~\ref{table:empirical}. Under the constant intervention effect specification, the estimated effects are $1.65$, $1.70$, and $1.85$ days under EXCH, NE, and DTD, respectively. All 95\% CIs exclude zero, suggesting that service removal is associated with a longer hospital stay under these models. The estimated time-averaged effects are larger, at $2.56$, $2.58$, and $2.84$ days, respectively, again with all 95\% CIs excluding zero. These findings are consistent with Kenny et al.,\cite{Kenny2022} which showed that assuming a constant intervention effect when the effect varies with exposure time can lead to underestimation of the time-averaged intervention effect. The estimated standard errors also depend on the working model and variance estimator. For instance, under the constant-effect EXCH model, the model-based standard error is $0.25$, compared with $0.31$ for the uncorrected RVE and $0.42$ for the MD-corrected RVE. Of note, the MD-corrected standard error is not uniformly larger than the model-based standard error. Under the exposure-time-dependent NE model, the corresponding values are $0.49$ and $0.58$.

We further examine sensitivity to including a random intervention effect under EXCH and NE (Web Table~15). The independent specification follows Section~\ref{sec:intervention}, whereas the correlated specification allows the random intervention effect to correlate with the cluster-level random intercept, relaxing the constraint that the within-period ICC under intervention is at least as large as that under control.\cite{Hemming2018b} Under EXCH, both specifications give a constant-effect estimate of $1.63$ days, while the estimated time-averaged effects are $2.64$ and $2.56$ days under the independent and correlated specifications, respectively. Under NE, the estimates and standard errors change little, although both specifications yield singular fits, indicating that the estimated random-effects covariance matrices are on or near the boundary of the parameter space. All 95\% CIs exclude zero, and our previous conclusions change little with inclusion of a random intervention effect.

\section{Discussion}\label{sec:discussion}

Hooper and Copas\cite{Hooper2019} emphasized that SW-CRTs with continuous recruitment require special consideration at both the design and analysis stages, including a clear distinction from discrete sampling. Although the CONSORT 2010 extension for SW-CRTs requires transparent reporting of recruitment mechanisms,\cite{Hemming2018a} the analytical implications of continuous recruitment have remained unclear. In light of this context, our study conveys several main messages. First, our simulation findings support the consistency of intervention effect estimators from discrete-time linear mixed models under continuous recruitment when the intervention effect structure is correctly specified and recruitment is independent of the intervention condition. We observe this robustness pattern across settings with discrete or continuous period effects, with or without a random intervention effect, and with constant or exposure-time-dependent intervention effects under a CTD correlation structure. However, intervention-dependent recruitment patterns can introduce bias, as illustrated in Simulation Study~VII. Second, even when the intervention effect estimator is consistent, valid inference depends on the choice of variance estimator. Model-based variance estimators underestimate the empirical variances in most settings, most severely under EXCH. In settings without a random intervention effect, the NE and DTD working structures may instead overestimate the empirical variances. The RVE with MD correction consistently achieves nominal coverage in the settings with negligible point-estimator bias, although this correction does not address bias arising from intervention-dependent recruitment. These findings support wider use of the MD-corrected RVE in SW-CRTs, particularly because 95.1\% of cross-sectional SW-CRTs use continuous recruitment,\cite{Nevins2024} while discrete-time linear mixed models remain common in practice. Third, Simulation Study~VIII evaluates the performance and computational demands of CTD working models. The CTD working model with discrete period indicators yields model-based 95\% CI coverage similar to that under NE and DTD across both configurations. With the implementation and settings used in this study, however, a CTD fit requires roughly an hour on average at $I = 32$, compared with less than a second for each discrete-time working model. These results demonstrate the computational advantages of discrete-time working models within the settings considered. Together with the earlier simulation findings, they support the practical use of discrete-time linear mixed models with the MD-corrected RVE, while emphasizing the need to assess whether recruitment patterns depend on the intervention condition.

An important exception to these generally reassuring findings arises when recruitment patterns change systematically between control and intervention periods. In Simulation Study VII, shifting from uniform recruitment under control to cluster-period mixed recruitment under intervention produces biased intervention effect estimates, even at large sample sizes. In practice, such recruitment patterns may be shaped by two possible mechanisms. The first is intervention-driven: growing awareness of interventions such as cash transfers or housing assistance may increase participant interest and referrals over time. The second is calendar-driven: recruitment may vary seasonally, for example, with reduced enrollment during holidays. Such variation may contribute to differences in recruitment timing between control and intervention observations within the same period. In both cases, the intervention alters who is recruited, so the treatment and control samples are no longer comparable. This is a form of selection bias,\cite{Li2022} and it is where the model-robustness of discrete-time linear mixed models breaks down: the bias lies in the estimand itself, beyond the reach of any working correlation structure or variance estimator. No method currently addresses recruitment-driven selection bias in the stepped wedge setting, and even for parallel-arm designs such methods are only beginning to emerge;\cite{Papadogeorgou2025} developing them is an important direction for future work. In the meantime, investigators can explore recruitment patterns before formal analysis using the interactive Shiny application we developed (\url{https://f07k8s-hao-wang.shinyapps.io/continuous-recruitment-swcrt/}), which visualizes continuous recruitment across clusters and periods. A tutorial is provided in Web Appendix~A.

Building on these findings, we offer several considerations for trial design and analysis. At the design stage, our simulations suggest that investigators can allow flexibility in recruitment timing when recruitment is independent of the intervention condition. Investigators should nonetheless establish protocols for documenting recruitment patterns over the course of the trial. Individual-level recruitment times can be collected systematically when feasible. When confidentiality concerns limit access to exact enrollment dates, aggregated summaries, such as cluster-period-specific enrollment densities, may still support descriptive assessment. At the analysis stage, investigators can first examine recruitment patterns under the control and intervention conditions using our interactive Shiny application. When recruitment is independent of the intervention condition and the intervention effect structure is correctly specified, our simulation findings support discrete-time analyses with the RVE and MD correction. However, systematic changes in recruitment between conditions may introduce non-negligible bias in the intervention effect estimator. Of note, in Simulation Study~VII, the estimated standard errors closely approximate the empirical standard deviations, suggesting that the resulting undercoverage is primarily caused by the bias in the point estimator. Moreover, the degree of undercoverage depends on the magnitude of this bias relative to the standard error. When the bias remains approximately constant as the sample size grows, smaller standard errors can lead to more severe undercoverage. For instance, in Simulation Study~VII, coverage of confidence intervals based on the RVE with MD correction decreases from approximately $95\%$ at $I = 16$ to approximately $90\%$ at $I = 96$. Thus, larger sample sizes and small-sample variance corrections do not resolve this source of bias. Further work is needed to develop and evaluate causal inference methods that target clearly defined estimands while accounting for intervention-dependent recruitment in SW-CRTs.

Although we investigate the impact of continuous recruitment on discrete-time linear mixed models across a wide range of settings, this study has several limitations. First, valid inference for the intervention effect does not ensure accurate estimation of the underlying covariance parameters. In particular, variance component estimates under the EXCH, NE, and DTD working structures may not adequately describe the CTD correlation structure in the data-generating process. An alternative is to fit a CTD working model directly using Markov chain Monte Carlo maximum likelihood,\cite{Samuel2024} although this approach was computationally demanding with the implementation and settings used in Simulation Study~VIII. For these CTD fits, we report model-based standard errors and do not evaluate robust variance estimation with small-sample corrections. We therefore encourage further work on efficient estimation of CTD models and the performance of robust variance estimators for these models. Such work is particularly important because ICC estimates from pilot studies or completed trials are routinely used for sample size calculations in future SW-CRTs. Reliable estimation of the variance and decay parameters under CTD, together with guidance for incorporating these parameters into design calculations for discrete-time analyses, would be useful when recruitment-time information is available. Second, we focus on continuous outcomes in this initial effort. Extending our simulation studies to binary outcomes is an important direction for future work. One caveat is that intervention effect estimates from logistic generalized linear mixed models can be sensitive to the specification of the random-effects structure. However, this limitation should not preclude investigating how continuous recruitment affects the performance of binary outcome analyses.

\bibliographystyle{SageV}
\bibliography{reference}

@article{Bell2002,
    author    = {Bell, Robert M. and McCaffrey, Daniel F.},
    title     = {Bias Reduction in Standard Errors for Linear Regression with Multi-Stage Samples},
    journal   = {Survey Methodology},
    volume    = {28},
    number    = {2},
    pages     = {169--181},
    year      = {2002},
    publisher = {Statistics Canada}
}

@article{Bowden2021,
    author = {Rhys Bowden and Andrew B Forbes and Jessica Kasza},
    title ={Inference for the treatment effect in longitudinal cluster randomized trials when treatment effect heterogeneity is ignored},
    journal = {Statistical Methods in Medical Research},
    volume = {30},
    number = {11},
    pages = {2503--2525},
    year = {2021},
}

@article{Copas2015,
    author = {Copas, AJ and Lewis, JJ and Thompson, JA and Davey, C and Baio, G and Hargreaves, JR},
    journal = {Trials},
    number = {1},
    publisher = {Trials},
    title = {Designing a stepped wedge trial: three main designs, carry-over effects and randomisation approaches},
    volume = {16},
    year = {2015}
}

@article{DavisPlourde2021,
    title = {Sample size considerations for stepped wedge designs with subclusters},
    volume = {79},
    ISSN = {1541-0420},
    number = {1},
    journal = {Biometrics},
    publisher = {Oxford University Press (OUP)},
    author = {Davis‐Plourde,  Kendra and Taljaard,  Monica and Li,  Fan},
    year = {2021},
    month = nov,
    pages = {98–-112}
}

@article{Easter2021,
    title = {Cluster randomized trials of individual-level interventions were at high risk of bias},
    journal = {Journal of Clinical Epidemiology},
    volume = {138},
    pages = {49--59},
    year = {2021},
    issn = {0895-4356},
    author = {Christina Easter and Jennifer A. Thompson and Sandra Eldridge and Monica Taljaard and Karla Hemming},
}

@article{Girling2016,
    author = {Girling, Alan J. and Hemming, Karla},
    title = {Statistical efficiency and optimal design for stepped cluster studies under linear mixed effects models},
    journal = {Statistics in Medicine},
    volume = {35},
    number = {13},
    pages = {2149--2166},
    year = {2016}
}

@article{Grantham2019,
    author = {Grantham, Kelsey L. and Kasza, Jessica and Heritier, Stephane and Hemming, Karla and Forbes, Andrew B.},
    title = {Accounting for a decaying correlation structure in cluster randomized trials with continuous recruitment},
    journal = {Statistics in Medicine},
    volume = {38},
    number = {11},
    pages = {1918--1934},
    year = {2019}
}

@article{Ford2020,
    author = {Ford, Whitney P. and Westgate, Philip M.},
    title = {Maintaining the validity of inference in small-sample stepped wedge cluster randomized trials with binary outcomes when using generalized estimating equations},
    journal = {Statistics in Medicine},
    volume = {39},
    number = {21},
    pages = {2779--2792},
    year = {2020}
}

@article{Hussey2007,
    title = {Design and analysis of stepped wedge cluster randomized trials},
    journal = {Contemporary Clinical Trials},
    volume = {28},
    number = {2},
    pages = {182--191},
    year = {2007},
    issn = {1551-7144},
    author = {Michael A. Hussey and James P. Hughes}
}

@article{Hughes2015,
    title = {Current issues in the design and analysis of stepped wedge trials},
    journal = {Contemporary Clinical Trials},
    volume = {45},
    pages = {55--60},
    year = {2015},
    note = {10th Anniversary Special Issue},
    issn = {1551-7144},
    author = {James P. Hughes and Tanya S. Granston and Patrick J. Heagerty}
}

@article{Hooper2016,
    author = {Hooper, Richard and Teerenstra, Steven and de Hoop, Esther and Eldridge, Sandra},
    title = {Sample size calculation for stepped wedge and other longitudinal cluster randomised trials},
    journal = {Statistics in Medicine},
    volume = {35},
    number = {26},
    pages = {4718--4728},
    year = {2016}
}

@article{Haines2017,
    author = {Haines, Terry P. AND Bowles, Kelly-Ann AND Mitchell, Deb AND O’Brien, Lisa AND Markham, Donna AND Plumb, Samantha AND May, Kerry AND Philip, Kathleen AND Haas, Romi AND Sarkies, Mitchell N. AND Ghaly, Marcelle AND Shackell, Melina AND Chiu, Timothy AND McPhail, Steven AND McDermott, Fiona AND Skinner, Elizabeth H.},
    journal = {PLOS Medicine},
    publisher = {Public Library of Science},
    title = {Impact of disinvestment from weekend allied health services across acute medical and surgical wards: 2 stepped-wedge cluster randomised controlled trials},
    year = {2017},
    month = {10},
    volume = {14},
    pages = {1--21},
    number = {10},

}

@article{Hemming2017,
    title = {Analysis of cluster randomised stepped wedge trials with repeated cross-sectional samples},
    volume = {18},
    ISSN = {1745--6215},
    number = {1},
    journal = {Trials},
    publisher = {Springer Science and Business Media LLC},
    author = {Hemming,  Karla and Taljaard,  Monica and Forbes,  Andrew},
    year = {2017},
    month = mar 
}

@article{Hemming2018a,
    author = {Hemming, Karla and Taljaard, Monica and McKenzie, Joanne E and Hooper, Richard and Copas, Andrew and Thompson, Jennifer A and Dixon-Woods, Mary and Aldcroft, Adrian and Doussau, Adelaide and Grayling, Michael and Kristunas, Caroline and Goldstein, Cory E and Campbell, Marion K and Girling, Alan and Eldridge, Sandra and Campbell, Mike J and Lilford, Richard J and Weijer, Charles and Forbes, Andrew B and Grimshaw, Jeremy M},
    title = {Reporting of stepped wedge cluster randomised trials: extension of the CONSORT 2010 statement with explanation and elaboration},
    volume = {363},
    elocation-id = {k1614},
    year = {2018},
    publisher = {BMJ Publishing Group Ltd},
    issn = {0959-8138},
    journal = {BMJ}
}

@article{Hemming2018b,
    author = {Hemming, Karla and Taljaard, Monica and Forbes, Andrew},
    title = {Modeling clustering and treatment effect heterogeneity in parallel and stepped-wedge cluster randomized trials},
    journal = {Statistics in Medicine},
    volume = {37},
    number = {6},
    year = {2018},
    pages = {883--898},
}

@article{Hooper2019,
    title = {Stepped wedge trials with continuous recruitment require new ways of thinking},
    journal = {Journal of Clinical Epidemiology},
    volume = {116},
    pages = {161--166},
    year = {2019},
    issn = {0895-4356},
    author = {Richard Hooper and Andrew Copas},
}

@article{Hooper2020,
    title = {The hunt for efficient, incomplete designs for stepped wedge trials with continuous recruitment and continuous outcome measures},
    volume = {20},
    ISSN = {1471-2288},
    number = {1},
    journal = {BMC Medical Research Methodology},
    publisher = {Springer Science and Business Media LLC},
    author = {Hooper,  Richard and Kasza,  Jessica and Forbes,  Andrew},
    year = {2020},
    month = Nov 
}

@article{Hooper2021,
    title = {Key concepts in clinical epidemiology: Stepped wedge trials},
    volume = {137},
    journal = {Journal of Clinical Epidemiology},
    publisher = {Elsevier BV},
    author = {Hooper, Richard},
    year = {2021},
    month = sep,
    pages = {159–-162}
}

@article{Hooper2024,
    author = {Richard Hooper and Olivier Quintin and Jessica Kasza},
    title ={Efficient designs for three-sequence stepped wedge trials with continuous recruitment},
    journal = {Clinical Trials},
    volume = {21},
    number = {6},
    pages = {723--733},
    year = {2024},
}

@article{Kasza2019a,
    author = {J Kasza and K Hemming and R Hooper and JNS Matthews and AB Forbes},
    title ={Impact of non-uniform correlation structure on sample size and power in multiple-period cluster randomised trials},
    journal = {Statistical Methods in Medical Research},
    volume = {28},
    number = {3},
    pages = {703--716},
    year = {2019},
}

@article{Kasza2019b,
    author = {Jessica Kasza and Andrew B Forbes},
    title ={Inference for the treatment effect in multiple-period cluster randomised trials when random effect correlation structure is misspecified},
    journal = {Statistical Methods in Medical Research},
    volume = {28},
    number = {10-11},
    pages = {3112--3122},
    year = {2019},
}

@article{Kasza2020,
    author = {Kasza, Jessica and Hooper, Richard and Copas, Andrew and Forbes, Andrew B.},
    title = {Sample size and power calculations for open cohort longitudinal cluster randomized trials},
    journal = {Statistics in Medicine},
    volume = {39},
    number = {13},
    pages = {1871--1883},
    year = {2020}
}

@article{Korevaar2021,
    author = {Elizabeth Korevaar and Jessica Kasza and Monica Taljaard and Karla Hemming and Terry Haines and Elizabeth L Turner and Jennifer A Thompson and James P Hughes and Andrew B Forbes},
    title ={Intra-cluster correlations from the CLustered OUtcome Dataset bank to inform the design of longitudinal cluster trials},
    journal = {Clinical Trials},
    volume = {18},
    number = {5},
    pages = {529--540},
    year = {2021},
}

@article{Kenny2022,
    author = {Kenny, A and Voldal, EC and Xia, F and Heagerty, PJ and Hughes, JP},
    title = {Analysis of stepped wedge cluster randomized trials in the presence of a time-varying treatment effect},
    journal = {Statistics in Medicine},
    volume = {41},
    number = {22},
    pages = {4311--4339},
    year = {2022}
}

@article{Kasza2022,
    author = {Kasza, Jessica and Bowden, Rhys and Hooper, Richard and Forbes, Andrew B.},
    title = {The batched stepped wedge design: A design robust to delays in cluster recruitment},
    journal = {Statistics in Medicine},
    volume = {41},
    number = {18},
    pages = {3627--3641},
    year = {2022}
}

@article{Liang1986,
    title = {Longitudinal data analysis using generalized linear models},
    volume = {73},
    ISSN = {1464-3510},
    number = {1},
    journal = {Biometrika},
    publisher = {Oxford University Press (OUP)},
    author = {Liang, Kung-Yee and Zeger,  Scott L.},
    year = {1986},
    pages = {13--22}
}

@article{Li2019,
    title = {Design and analysis considerations for cohort stepped wedge cluster randomized trials with a decay correlation structure},
    volume = {39},
    ISSN = {1097-0258},
    number = {4},
    journal = {Statistics in Medicine},
    publisher = {Wiley},
    author = {Li,  Fan},
    year = {2019},
    month = dec,
    pages = {438–-455}
}

@article{Li2020,
    author = {Fan Li and James P Hughes and Karla Hemming and Monica Taljaard and Edward R. Melnick and Patrick J Heagerty},
    title ={Mixed-effects models for the design and analysis of stepped wedge cluster randomized trials: an overview},
    journal = {Statistical Methods in Medical Research},
    volume = {30},
    number = {2},
    pages = {612--639},
    year = {2021},
}

@article{Li2021,
    title = {Marginal modeling of cluster-period means and intraclass correlations in stepped wedge designs with binary outcomes},
    volume = {23},
    ISSN = {1468-4357},
    number = {3},
    journal = {Biostatistics},
    publisher = {Oxford University Press (OUP)},
    author = {Li,  Fan and Yu,  Hengshi and Rathouz,  Paul J and Turner,  Elizabeth L and Preisser,  John S},
    year = {2021},
    month = feb,
    pages = {772–-788}
}

@article{Li2022,
    author = {Fan Li and Zizhong Tian and Jennifer Bobb and Georgia Papadogeorgou and Fan Li},
    title ={Clarifying selection bias in cluster randomized trials},
    journal = {Clinical Trials},
    volume = {19},
    number = {1},
    pages = {33--41},
    year = {2022},
}

@article{Lee2025,
    author = {Lee, Kenneth M. and Turner, Elizabeth L. and Kenny, Avi},
    title = {Analysis of Stepped-Wedge Cluster Randomized Trials When Treatment Effects Vary by Exposure Time or Calendar Time},
    journal = {Statistics in Medicine},
    volume = {44},
    number = {20-22},
    pages = {e70256},
    year = {2025}
}

@Article{Mancl2001,
    author={Lloyd A. Mancl and Timothy A. DeRouen},
    title={{A Covariance Estimator for GEE with Improved Small‐Sample Properties}},
    journal={Biometrics},
    year={2001},
    volume={57},
    number={1},
    pages={126--134},
    month={March},
}

@book{Murray2007,
    title = {Design and Analysis of Group-Randomized Trials},
    publisher = {Oxford University Press},
    author = {Murray,  David M},
    year = {2007},
    month = aug 
}

@article{Morris2019,
    author = {Morris, Tim P. and White, Ian R. and Crowther, Michael J.},
    title = {Using simulation studies to evaluate statistical methods},
    journal = {Statistics in Medicine},
    volume = {38},
    number = {11},
    pages = {2074--2102},
    year = {2019}
}

@article{Maleyeff2022,
    author = {Maleyeff, L and Li, F and Haneuse, S and Wang, R},
    title = {Assessing exposure-time treatment effect heterogeneity in stepped-wedge cluster randomized trials},
    journal = {Biometrics},
    volume = {79},
    number = {3},
    pages = {2551--2564},
    year = {2022},
    month = {11},
    issn = {0006-341X},
}

@article{Nevins2023,
    title = {A scoping review described diversity in methods of randomization and reporting of baseline balance in stepped-wedge cluster randomized trials},
    journal = {Journal of Clinical Epidemiology},
    volume = {157},
    pages = {134--145},
    year = {2023},
    issn = {0895-4356},
    author = {Pascale Nevins and Kendra Davis-Plourde and Jules Antoine {Pereira Macedo} and Yongdong Ouyang and Mary Ryan and Guangyu Tong and Xueqi Wang and Can Meng and Luis Ortiz-Reyes and Fan Li and Agnès Caille and Monica Taljaard},
}

@article{Nevins2024,
    author = {Pascale Nevins and Mary Ryan and Kendra Davis-Plourde and Yongdong Ouyang and Jules Antoine Pereira Macedo and Can Meng and Guangyu Tong and Xueqi Wang and Luis Ortiz-Reyes and Agnès Caille and Fan Li and Monica Taljaard},
    title ={Adherence to key recommendations for design and analysis of stepped-wedge cluster randomized trials: A review of trials published 2016–2022},
    journal = {Clinical Trials},
    volume = {21},
    number = {2},
    pages = {199--210},
    year = {2024},
}

@article{Ouyang2024,
    author = {Yongdong Ouyang and Monica Taljaard and Andrew B Forbes and Fan Li},
    title ={Maintaining the validity of inference from linear mixed models in stepped-wedge cluster randomized trials under misspecified random-effects structures},
    journal = {Statistical Methods in Medical Research},
    volume = {33},
    number = {9},
    pages = {1497--1516},
    year = {2024},
}

@article{Papadogeorgou2025,
    author = {Papadogeorgou, Georgia and Liu, Bo and Li, Fan and Li, Fan},
    title = {Addressing selection bias in cluster randomized experiments via weighting},
    journal = {Biometrics},
    volume = {81},
    number = {1},
    pages = {ujaf013},
    year = {2025},
    month = {03},
    issn = {0006-341X},
}

@article{Samuel2024,
      title={Generalised Linear Mixed Model Specification, Analysis, Fitting, and Optimal Design in R with the glmmr Packages}, 
      author={Samuel I. Watson},
      year={2024},
      journal={arXiv preprint arXiv:2303.12657},
}

@article{Turner2017,
    author = {Turner, Elizabeth L. and Prague, Melanie and Gallis, John A. and Li, Fan and Murray, David M.},
    title = {Review of Recent Methodological Developments in Group-Randomized Trials: Part 2—Analysis},
    journal = {American Journal of Public Health},
    volume = {107},
    number = {7},
    pages = {1078--1086},
    year = {2017},
}

@article{Taljaard2020,
    author = {Monica Taljaard and Cory E Goldstein and Bruno Giraudeau and Stuart G Nicholls and Kelly Carroll and Spencer Phillips Hey and Jamie C Brehaut and Vipul Jairath and Alex John London and Sandra M Eldridge and Jeremy M Grimshaw and Dean A Fergusson and Charles Weijer},
    title ={Cluster over individual randomization: are study design choices appropriately justified? Review of a random sample of trials},
    journal = {Clinical Trials},
    volume = {17},
    number = {3},
    pages = {253--263},
    year = {2020},
}

@article{Wang2024,
    author = {Wang, B and Wang, X and Li, F},
    title = {How to achieve model-robust inference in stepped wedge trials with model-based methods?},
    journal = {Biometrics},
    volume = {80},
    number = {4},
    pages = {ujae123},
    year = {2024},
    month = {11},
    issn = {0006-341X},
}

@article{Wang2026,
    author = {Wang, Hao and Chen, Xinyuan and Courtright, Katherine R. and Halpern, Scott D. and Harhay, Michael O. and Taljaard, Monica and Li, Fan},
    title = {On Anticipation Effect in Stepped Wedge Cluster Randomized Trials},
    journal = {Statistics in Medicine},
    volume = {45},
    number = {3-5},
    pages = {e70380},
    year = {2026}
}

@article{Voldal2022,
    author = {Voldal, Emily C. and Xia, Fan and Kenny, Avi and Heagerty, Patrick J. and Hughes, James P.},
    title = {Model misspecification in stepped wedge trials: Random effects for time or treatment},
    journal = {Statistics in Medicine},
    volume = {41},
    number = {10},
    pages = {1751--1766},
    year = {2022}
}

\end{document}


\runninghead{}

\renewcommand{\figurename}{Web Figure}
\setcounter{figure}{0}
\renewcommand{\tablename}{Web Table}
\setcounter{table}{0}

\begin{center}
\textbf{\Large
Web Appendix for \\ ``Can discrete-time analyses be trusted for stepped wedge trials with continuous recruitment?''}\\
\vspace{1em}
    {\normalsize Hao Wang, Guangyu Tong, Heather Allore, Kelsey L. Grantham, Monica Taljaard, and Fan Li$^{\ast}$\par}
\end{center}

\section*{Web Appendix A: Tutorial for the Continuous Recruitment Pattern Visualization App}\label{sec:web_appendix_A}

In this appendix, we describe an interactive Shiny application for visualizing continuous recruitment patterns across clusters and periods in SW-CRTs. The application allows investigators to assess systematic differences in recruitment patterns between control and intervention periods before formal analysis. The application is publicly available at \href{https://f07k8s-hao-wang.shinyapps.io/continuous-recruitment-swcrt/}{https://f07k8s-hao-wang.shinyapps.io/continuous-recruitment-swcrt/}, and the source code is available at \href{https://github.com/haowangbiostat/Continuous-recruitment-SWCRT}{https://github.com/haowangbiostat/Continuous-recruitment-SWCRT}.

\subsection*{A.1 Overview}

The application allows investigators to (i) upload individual-level recruitment data from their own trials, (ii) visualize recruitment patterns separately for each cluster, and (iii) compare patterns under the control and intervention conditions within each cluster. An illustrative example dataset is loaded by default so that investigators can explore the application before uploading their own data.

\subsection*{A.2 Data Format and Requirements}

To use the application with their own data, investigators upload a comma-separated values (CSV) file with one row per individual and the following four columns:
\begin{itemize}
    \item \texttt{id\_cluster}: a unique identifier for each cluster.
    \item \texttt{time}: the discrete period index $j \in \{1, \ldots, J\}$ indicating when the individual was recruited.
    \item \texttt{cal}: the calendar date of recruitment in \texttt{YYYY-MM-DD} format.
    \item \texttt{trt}: a binary treatment indicator, equal to $0$ under control and $1$ under intervention.
\end{itemize}
The application standardizes calendar dates to express recruitment times in units of periods, using the time scale introduced in Section~2.1 of the main paper. Specifically, for each cluster, the application (i) uses the earliest observed recruitment date as the reference date, (ii) calculates the number of days elapsed from this date for each individual, and (iii) divides the elapsed time by the empirical period length.

\subsection*{A.3 Instructions}

\begin{figure}[!htbp]
    \centering
    \includegraphics[width=\linewidth]{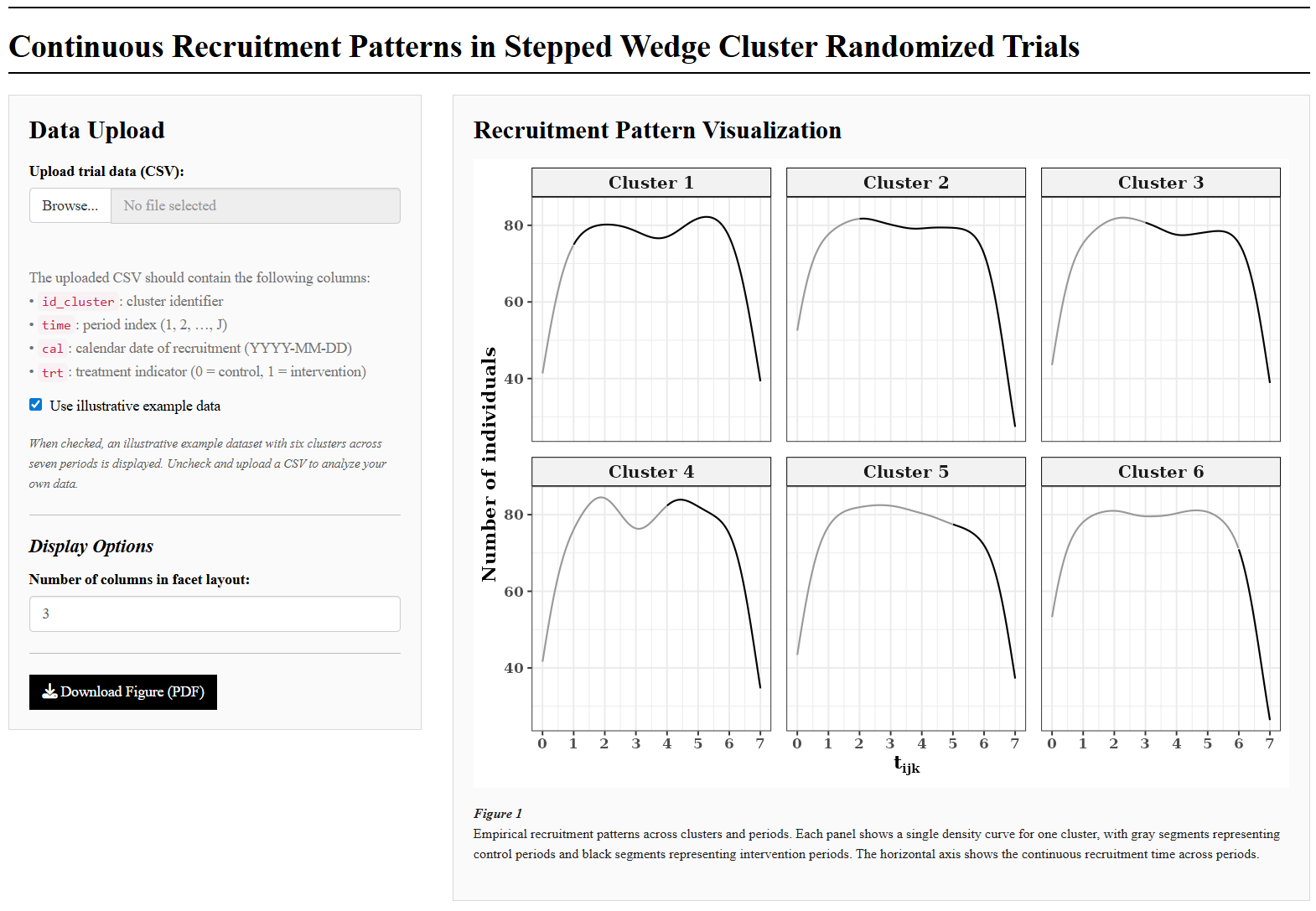}
    \caption{Screenshot of the interactive Shiny application for visualizing continuous recruitment patterns in SW-CRTs. The left panel provides a data upload interface, a toggle for displaying the illustrative example dataset, a display option for adjusting the number of columns in the facet layout, and a download button for exporting the resulting figure. The right panel presents the recruitment pattern visualization, where each subpanel displays a single density curve for one cluster across periods, with gray segments representing control periods and black segments representing intervention periods. The horizontal axis shows the continuous recruitment time $t_{ijk}$ across periods.}
    \label{fig:shiny_app}
\end{figure}

Using the application involves three steps. First, investigators open the application URL in a web browser. The application automatically loads an illustrative example dataset with six clusters and seven periods (Web Figure~\ref{fig:shiny_app}), allowing investigators to explore the interface and visualization before uploading their own data.

Second, investigators uncheck the ``Use illustrative example data'' option and upload their CSV file to visualize their own trial data. The application checks the required columns and computes the standardized recruitment time $t_{ijk}$ for each individual. It then generates a recruitment density curve in a separate panel for each cluster. The horizontal axis represents $t_{ijk}$, and the vertical axis represents the estimated recruitment intensity, expressed as the number of individuals recruited per unit of time. The curve is colored gray under control and black under intervention, with the color transition determined by the earliest observed recruitment time under intervention.

Third, investigators examine whether recruitment patterns differ systematically between control and intervention periods within each cluster. Potential indicators of intervention-dependent recruitment include (i) a substantial change in the shape of the curve between the gray and black segments, (ii) a marked increase or decrease in recruitment intensity near the transition to intervention, or (iii) different within-period recruitment patterns, such as approximately uniform recruitment under control and concentrated recruitment under intervention. These patterns warrant closer examination of the recruitment process and its potential implications for bias in the intervention effect estimates. Investigators can also adjust the number of columns used to arrange the cluster panels and download the resulting figure in PDF format for inclusion in reports or manuscripts.

\clearpage
\section*{Web Appendix B: Additional Figures and Tables}

\begin{table}[t]
\caption{Parameter values used in the simulation studies under the CTD correlation structure, with and without an RI effect. Parameter combinations are study-specific.}\label{table:set_up}
\resizebox{\linewidth}{!}{
    \begin{tabular}{ll}
        \toprule
        \textbf{Parameter} & \textbf{Values} \\
        \midrule
        Number of clusters ($I$) & $\{16, 32, 96\}$ \\
        Number of periods ($J$) & $\{5, 9\}$ \\
        Number of sequences ($Q = J-1$) & $\{4, 8\}$ \\
        Equal cluster-period size ($K$) & $50$ by default, with $20$ in Simulation Study~VIII when $I = 16$ \\
        Intervention-dependent recruitment size ($K_{ij}$) & Balanced: $(50, 50)$; moderately unbalanced: $(25, 75)$; severely unbalanced: $(10, 90)$ \\
        Exposure-time-dependent recruitment size ($K_{ij}$) & S1: $(25, 50, 75, 100, 125)$; S2: $(10, 100, 110, 130, 160)$ \\
        Continuous recruitment pattern & Uniform, cluster mixed, cluster-period mixed \\
        Recruitment patterns under control and intervention & U + U, U + CP, CP + CP in Simulation Study~VII \\
        Discrete period effect & $\beta_j = 0.5j^2/J$ \\
        Continuous period effect & $T(t_{ijk}) = \{0.5t_{ijk}^{2}+\sin(6\pi t_{ijk})\}/J$ \\
        Constant intervention effect ($\delta$) & $0$, with $2$ in one additional setting in Simulation Study~VII \\
        Time-averaged intervention effect ($\Delta$) & $0$ \\
        Point intervention effects $(\delta(1),\ldots,\delta(4))$ & Approximately $(-1.38, 0.38, 0.98, 0.02)$ \\
        \midrule
        \multicolumn{2}{l}{\textit{With an RI effect}} \\
        \hspace*{1em}ICC at zero time separation under control ($\rho_0$) & $\{0.01, 0.05\}$ \\
        \hspace*{1em}ICC at zero time separation under intervention ($\rho_1$) & $0.1$ \\
        \hspace*{1em}CAC at a separation of one period under control ($r$) & $\{0.5, 0.8\}$ \\
        \midrule
        \multicolumn{2}{l}{\textit{Without an RI effect}} \\
        \hspace*{1em}ICC at zero time separation ($\rho_0 = \rho_1$) & $\{0.01, 0.05\}$ \\
        \hspace*{1em}CAC at a separation of one period ($r$) & $\{0.5, 0.8\}$ \\
        \midrule
        Residual variance ($\sigma_\epsilon^2$) & $1$ \\
        \bottomrule
    \end{tabular}
}\par\vspace{5pt}
{\scriptsize EXCH: simple exchangeable; NE: nested exchangeable; DTD: discrete-time exponential decay; CTD: continuous-time exponential decay; RI: random intervention; ICC: intracluster correlation coefficient; CAC: cluster autocorrelation coefficient. For intervention-dependent recruitment size, the two entries denote the number of individuals per cluster-period under control and intervention, respectively. For exposure-time-dependent recruitment size, the five entries denote the number of individuals per cluster-period under control and at exposure times $1, 2, 3, 4$, respectively. U and CP denote uniform and cluster-period mixed recruitment, respectively, with the entries before and after ``+'' corresponding to control and intervention. The additional setting with $\delta = 2$ is restricted to Simulation Study~VII under $I = 96$, S2, and U + CP.\par}
\end{table}

\begin{table}[t]
    \caption{Summary of performance measures for a constant intervention effect. Here, $n$ denotes the number of simulation replicates included in each calculation, and $\bar\delta = n^{-1}\sum_{p=1}^n\hat\delta_p$ is the empirical mean of the intervention effect estimates. The quantities $\hat\delta_{p,\text{low}}$ and $\hat\delta_{p,\text{high}}$ denote the lower and upper limits of the 95\% CI in replicate $p$. For an exposure-time-dependent intervention effect, the same definitions apply to each point effect $\delta(s)$ and the time-averaged effect $\Delta$, using their corresponding estimates and confidence intervals.}\label{table:metrics}
    \centering
    \resizebox{0.8\linewidth}{!}{
    \begin{tabular}{lll}
        \toprule
        \textbf{Measure} & \textbf{Target Quantity} & \textbf{Monte Carlo Estimate} \\
        \midrule
        Bias &
        $\bbE(\hat\delta)-\delta$ &
        $\displaystyle \frac{1}{n}\sum_{p=1}^n(\hat\delta_p-\delta)$ \\[10pt]

        95\% CI coverage (\%) &
        $100\Pr(\hat\delta_{\text{low}}\leq\delta\leq\hat\delta_{\text{high}})$ &
        $\displaystyle \frac{100}{n}\sum_{p=1}^n
        \bbone(\hat\delta_{p,\text{low}}\leq\delta\leq\hat\delta_{p,\text{high}})$ \\[10pt]

        Average estimated standard error\textsuperscript{\textdagger} &
        $\displaystyle \bbE\left\{\sqrt{\widehat{\bbV}(\hat\delta)}\right\}$ &
        $\displaystyle \frac{1}{n}\sum_{p=1}^n
        \sqrt{\widehat{\bbV}(\hat\delta_p)}$ \\[10pt]

        Empirical standard deviation &
        $\displaystyle \sqrt{\bbV(\hat\delta)}$ &
        $\displaystyle \sqrt{\frac{1}{n-1}
        \sum_{p=1}^n(\hat\delta_p-\bar\delta)^2}$ \\[10pt]
        \bottomrule
    \end{tabular}
    }\par\vspace{5pt}
    {\scriptsize\raggedright
    \textsuperscript{\textdagger}:
    $\widehat{\bbV}(\hat\delta_p)$ denotes the estimated variance in replicate $p$, obtained using the model-based estimator $\bbV_{\text{Naive}}(\hat\delta_p)$, the uncorrected robust variance estimator $\bbV_{\text{RVE}}(\hat\delta_p)$, or the MD-corrected robust variance estimator $\bbV_{\text{RVE}}^{\text{MD}}(\hat\delta_p)$. In contrast, $\bbV(\hat\delta)$ denotes the sampling variance of the intervention effect estimator. CI: confidence interval; RVE: robust variance estimator; MD: Mancl and DeRouen.\par}
\end{table}

\begin{table}[h]
    \small
    \centering
    \caption{Overview of the simulation studies, with the scenario(s) of the true data-generating process considered and the corresponding research question for each study.}
    \resizebox{\linewidth}{!}{\begin{tabular}{llp{9.5cm}}
        \toprule
        Study & Scenario(s) & Research Question \\
        \midrule
        I (Section 3.5.1) & 1 versus 2 & How does the specification of the period effect (discrete vs.\ continuous) in the true data-generating process impact the performance of the discrete-time working models? \\
        II (Section 3.5.2) & 2 & How do different continuous recruitment patterns in the true data-generating process impact the performance of the discrete-time working models? \\
        III (Section 3.5.3) & 2 & How do varying ICC-related parameters in the true data-generating process impact the performance of the discrete-time working models? \\
        IV (Section 3.5.4) & 2 versus 3 & What is the impact of a random intervention effect in the true data-generating process on the performance of the discrete-time working models? \\
        V (Section 3.5.5) & 2 & How do trial design parameters (i.e., number of clusters and periods) in the true data-generating process impact the performance of the discrete-time working models? \\
        VI (Section 3.5.6) & 4 & What is the impact of an exposure-time-dependent intervention effect in the true data-generating process on the performance of the discrete-time working models? \\
        VII (Section 3.5.7) & 2 and 4 & How does intervention-dependent recruitment in the true data-generating process impact the performance of the discrete-time working models? \\
        VIII (Section 3.5.8) & 5 & How do the three discrete-time working models compare with the CTD working model? \\
        \bottomrule
    \end{tabular}}
    \label{tab:sim_overview}
\end{table}

\begin{table}[!htbp]
    \caption{Simulation results based on 2,000 datasets generated under Scenario 3 for Simulation Study IV. The design assumes a standard SW-CRT with $I = 32$, $J = 5$, $K_{ij} = 50$, $\delta = 0$, $\rho_0 \in \{0.01, 0.05\}$, $\sigma_\epsilon^{2} = 1$, CAC $\in \{0.5, 0.8\}$, a continuous period effect $T(t_{ijk}) = \{0.5 t_{ijk}^{2} + \sin(6\pi t_{ijk})\}/J$, a cluster-period mixed recruitment pattern, and no random intervention effect. $\sd(\cdot)$ denotes the empirical (Monte Carlo) standard deviation. $\se_{\text{Naive}}(\cdot)$, $\se_{\text{RVE}}(\cdot)$, and $\se_{\text{RVE}}^{\text{MD}}(\cdot)$ denote the average model-based, robust, and MD-corrected robust standard errors, respectively. $\bbC_{\text{Naive}}(\cdot)$, $\bbC_{\text{RVE}}(\cdot)$, and $\bbC_{\text{RVE}}^{\text{MD}}(\cdot)$ denote the corresponding empirical coverage of the 95\% confidence intervals.}\label{tab:no_RI}
    \resizebox{\linewidth}{!}{
    \begin{tabular}{cclccccccccc}
        \toprule
        CAC & $\rho_0$ & Model & Bias & sd$(\hat\delta)$ & $\se_{\text{Naive}}(\hat\delta)$ & $\bbC_{\text{Naive}}(\hat\delta)$ & $\se_{\text{RVE}}(\hat\delta)$ & $\bbC_{\text{RVE}}(\hat\delta)$ & $\se_{\text{RVE}}^{\text{MD}}(\hat\delta)$ & $\bbC_{\text{RVE}}^{\text{MD}}(\hat\delta)$ & Convergence \\ 
        \midrule
        \multirow{6}{*}{0.5} & \multirow{3}{*}{0.01} & EXCH & 0.000 & 0.042 & 0.036 & 91.7 & 0.040 & 93.7 & 0.043 & 95.1 & 100 \\
        & & NE & 0.000 & 0.042 & 0.041 & 95.5 & 0.039 & 93.8 & 0.042 & 94.7 & 98 \\ 
        & & DTD & 0.000 & 0.042 & 0.042 & 96.0 & 0.039 & 93.8 & 0.042 & 95.0 & 99 \\ 
        \cmidrule{2-12}
        & \multirow{3}{*}{0.05} & EXCH & 0.001 & 0.061 & 0.040 & 81.8 & 0.058 & 93.1 & 0.062 & 94.5 & 100 \\ 
        & & NE & 0.001 & 0.060 & 0.058 & 94.8 & 0.057 & 93.2 & 0.061 & 95.2 & 99 \\
        & & DTD & 0.001 & 0.058 & 0.058 & 95.8 & 0.055 & 93.5 & 0.059 & 95.3 & 100 \\
        \midrule
        \multirow{6}{*}{0.8} & \multirow{3}{*}{0.01} & EXCH & 0.000 & 0.041 & 0.037 & 93.5 & 0.039 & 93.8 & 0.042 & 95.5 & 100 \\ 
        & & NE & 0.000 & 0.041 & 0.041 & 96.0 & 0.039 & 93.9 & 0.042 & 95.5 & 99 \\ 
        & & DTD & 0.000 & 0.041 & 0.042 & 95.7 & 0.039 & 93.8 & 0.042 & 95.4 & 97 \\
        \cmidrule{2-12}
        & \multirow{3}{*}{0.05} & EXCH & 0.000 & 0.053 & 0.040 & 88.0 & 0.051 & 94.2 & 0.054 & 95.6 & 100 \\
        & & NE & 0.000 & 0.054 & 0.054 & 96.0 & 0.051 & 93.3 & 0.054 & 95.6 & 99 \\
        & & DTD & -0.000 & 0.052 & 0.052 & 96.3 & 0.049 & 94.5 & 0.052 & 96.3 & 100 \\ 
        \bottomrule
    \end{tabular}
    }\par\vspace{5pt}
    {\scriptsize EXCH: simple exchangeable; NE: nested exchangeable; DTD: discrete-time exponential decay; ICC: intracluster correlation coefficients; CAC: cluster autocorrelation coefficient; $\rho_0$: within‑period ICC; RVE: robust variance estimator; MD: Mancl and DeRouen.\par}
\end{table}

\begin{table}[htbp]
    \caption{Simulation results based on 2,000 datasets generated under Scenario 2 for Simulation Study V. The design assumes a standard SW-CRT with $I = 16$, $J = 5$, $K_{ij} = 50$, $\delta = 0$, $\rho_0 \in \{0.01, 0.05\}$, $\rho_1 = 0.1$, $\sigma_\epsilon^{2} = 1$, CAC $\in \{0.5, 0.8\}$, a continuous period effect $T(t_{ijk}) = \{0.5 t_{ijk}^{2} + \sin(6\pi t_{ijk})\}/J$, and a cluster-period mixed recruitment pattern. $\sd(\cdot)$ denotes the empirical (Monte Carlo) standard deviation. $\se_{\text{Naive}}(\cdot)$, $\se_{\text{RVE}}(\cdot)$, and $\se_{\text{RVE}}^{\text{MD}}(\cdot)$ denote the average model-based, robust, and MD-corrected robust standard errors, respectively. $\bbC_{\text{Naive}}(\cdot)$, $\bbC_{\text{RVE}}(\cdot)$, and $\bbC_{\text{RVE}}^{\text{MD}}(\cdot)$ denote the corresponding empirical coverage of the 95\% confidence intervals.}\label{tab:number_of_clusters}
    \resizebox{\linewidth}{!}{
    \begin{tabular}{cclccccccccc}
        \toprule
        CAC & $\rho_0$ & Model & Bias & sd$(\hat\delta)$ & $\se_{\text{Naive}}(\hat\delta)$ & $\bbC_{\text{Naive}}(\hat\delta)$ & $\se_{\text{RVE}}(\hat\delta)$ & $\bbC_{\text{RVE}}(\hat\delta)$ & $\se_{\text{RVE}}^{\text{MD}}(\hat\delta)$ & $\bbC_{\text{RVE}}^{\text{MD}}(\hat\delta)$ & Convergence \\ 
        \midrule
        \multirow{6}{*}{0.5} & \multirow{3}{*}{0.01} & EXCH & 0.002 & 0.103 & 0.057 & 76.0 & 0.093 & 93.4 & 0.107 & 95.8 & 100 \\
        & & NE & 0.003 & 0.104 & 0.086 & 92.0 & 0.093 & 93.0 & 0.106 & 95.8 & 99 \\
        & & DTD & 0.002 & 0.104 & 0.083 & 90.9 & 0.092 & 93.0 & 0.106 & 95.9 & 100  \\
        \cmidrule{2-12}
         & \multirow{3}{*}{0.05} & EXCH & -0.001 & 0.108 & 0.058 & 75.0 & 0.096 & 93.8 & 0.110 & 96.5 & 100 \\ 
        & & NE & -0.001 & 0.108 & 0.094 & 94.0 & 0.095 & 93.0 & 0.109 & 96.3 & 100 \\ 
        & & DTD & 0.000 & 0.105 & 0.091 & 93.7 & 0.092 & 93.0 & 0.106 & 95.7 & 100 \\
        \midrule
        \multirow{6}{*}{0.8} & \multirow{3}{*}{0.01} & EXCH & 0.001 & 0.104 & 0.057 & 76.8 & 0.092 & 93.5 & 0.105 & 96.0 & 100 \\
        & & NE & 0.000 & 0.104 & 0.085 & 91.7 & 0.091 & 92.8 & 0.104 & 95.5 & 99 \\ 
        & & DTD & 0.000 & 0.104 & 0.081 & 90.5 & 0.091 & 92.5 & 0.104 & 95.4 & 100 \\
        \cmidrule{2-12}
         & \multirow{3}{*}{0.05} & EXCH & 0.001 & 0.098 & 0.058 & 80.0 & 0.088 & 92.8 & 0.100 & 95.7 & 100 \\
        & & NE & 0.000 & 0.098 & 0.088 & 94.5 & 0.088 & 93.0 & 0.100 & 95.7 & 99 \\ 
        & & DTD & 0.000 & 0.096 & 0.083 & 93.8 & 0.086 & 92.6 & 0.098 & 95.9 & 100 \\
        \bottomrule
    \end{tabular}
    }\par\vspace{5pt}
    {\scriptsize EXCH: simple exchangeable; NE: nested exchangeable; DTD: discrete-time exponential decay; ICC: intracluster correlation coefficients; CAC: cluster autocorrelation coefficient; $\rho_0$: within‑period ICC under control; $\rho_1$: within‑period ICC under treatment; RVE: robust variance estimator; MD: Mancl and DeRouen.\par}
\end{table}

\begin{table}[htbp]
    \caption{Simulation results based on 2,000 datasets generated under Scenario 2 for Simulation Study V. The design assumes a standard SW-CRT with $I = 32$, $J = 9$, $K_{ij} = 50$, $\delta = 0$, $\rho_0 \in \{0.01, 0.05\}$, $\rho_1 = 0.1$, $\sigma_\epsilon^{2} = 1$, CAC $\in \{0.5, 0.8\}$, a continuous period effect $T(t_{ijk}) = \{0.5 t_{ijk}^{2} + \sin(6\pi t_{ijk})\}/J$, and a cluster-period mixed recruitment pattern. $\sd(\cdot)$ denotes the empirical (Monte Carlo) standard deviation. $\se_{\text{Naive}}(\cdot)$, $\se_{\text{RVE}}(\cdot)$, and $\se_{\text{RVE}}^{\text{MD}}(\cdot)$ denote the average model-based, robust, and MD-corrected robust standard errors, respectively. $\bbC_{\text{Naive}}(\cdot)$, $\bbC_{\text{RVE}}(\cdot)$, and $\bbC_{\text{RVE}}^{\text{MD}}(\cdot)$ denote the corresponding empirical coverage of the 95\% confidence intervals.}\label{tab:number_of_periods}
    \resizebox{\linewidth}{!}{
    \begin{tabular}{cclccccccccc}
        \toprule
        CAC & $\rho_0$ & Model & Bias & sd$(\hat\delta)$ & $\se_{\text{Naive}}(\hat\delta)$ & $\bbC_{\text{Naive}}(\hat\delta)$ & $\se_{\text{RVE}}(\hat\delta)$ & $\bbC_{\text{RVE}}(\hat\delta)$ & $\se_{\text{RVE}}^{\text{MD}}(\hat\delta)$ & $\bbC_{\text{RVE}}^{\text{MD}}(\hat\delta)$ & Convergence \\ 
        \midrule
        \multirow{6}{*}{0.5} & \multirow{3}{*}{0.01} & EXCH & -0.003 & 0.070 & 0.029 & 60.8 & 0.065 & 93.5 & 0.070 & 95.0 & 100 \\
        & & NE & -0.003 & 0.069 & 0.044 & 80.0 & 0.064 & 93.5 & 0.069 & 95.3 & 93 \\
        & & DTD & -0.003 & 0.069 & 0.046 & 82.4 & 0.064 & 93.5 & 0.068 & 95.0 & 100 \\
        \cmidrule{2-12}
         & \multirow{3}{*}{0.05} & EXCH & 0.001 & 0.073 & 0.029 & 60.5 & 0.067 & 93.2 & 0.072 & 95.2 & 100 \\ 
        & & NE & 0.001 & 0.071 & 0.050 & 85.2 & 0.065 & 92.5 & 0.069 & 94.7 & 95 \\ 
        & & DTD & 0.001 & 0.068 & 0.053 & 89.0 & 0.062 & 93.1 & 0.066 & 94.7 & 100 \\ 
        \midrule
        \multirow{6}{*}{0.8} & \multirow{3}{*}{0.01} & EXCH & 0.001 & 0.071 & 0.029 & 59.9 & 0.065 & 93.0 & 0.069 & 95.0 & 100 \\ 
        & & NE & 0.001 & 0.070 & 0.044 & 79.5 & 0.064 & 93.0 & 0.068 & 94.7 & 92 \\ 
        & & DTD & 0.001 & 0.070 & 0.044 & 80.6 & 0.064 & 92.8 & 0.068 & 94.1 & 100 \\ 
        \cmidrule{2-12}
         & \multirow{3}{*}{0.05} & EXCH & -0.000 & 0.067 & 0.030 & 62.9 & 0.063 & 93.8 & 0.067 & 95.7 & 100 \\ 
        & & NE & -0.001 & 0.066 & 0.048 & 85.7 & 0.062 & 94.1 & 0.066 & 95.5 & 94 \\  
        & & DTD & -0.001 & 0.062 & 0.048 & 88.8 & 0.058 & 93.4 & 0.062 & 94.8 & 100 \\
        \bottomrule
    \end{tabular}
    }\par\vspace{5pt}
    {\scriptsize EXCH: simple exchangeable; NE: nested exchangeable; DTD: discrete-time exponential decay; ICC: intracluster correlation coefficients; CAC: cluster autocorrelation coefficient; $\rho_0$: within‑period ICC under control; $\rho_1$: within‑period ICC under treatment; RVE: robust variance estimator; MD: Mancl and DeRouen.\par}
\end{table}

\begin{table}[htbp]
    \caption{Simulation results based on 2,000 datasets generated under Scenario 4 for Simulation Study VI. The design assumes a standard SW-CRT with $I = 32$, $J = 5$, $K_{ij} = 50$, $\delta(1) \approx -1.3768$, $\delta(2) \approx 0.3831$, $\delta(3) \approx 0.9785$, $\delta(4) \approx 0.0152$, $\Delta = 0$, $\rho_0 = 0.01$, $\rho_1 = 0.1$, $\sigma_\epsilon^{2} = 1$, CAC $= 0.5$, a continuous period effect $T(t_{ijk}) = \{0.5 t_{ijk}^{2} + \sin(6\pi t_{ijk})\}/J$, and a cluster-period mixed recruitment pattern. $\sd(\cdot)$ denotes the empirical (Monte Carlo) standard deviation. $\se_{\text{Naive}}(\cdot)$, $\se_{\text{RVE}}(\cdot)$, and $\se_{\text{RVE}}^{\text{MD}}(\cdot)$ denote the average model-based, robust, and MD-corrected robust standard errors, respectively. $\bbC_{\text{Naive}}(\cdot)$, $\bbC_{\text{RVE}}(\cdot)$, and $\bbC_{\text{RVE}}^{\text{MD}}(\cdot)$ denote the corresponding empirical coverage of the 95\% confidence intervals.}\label{tab:TVE}
    \resizebox{\linewidth}{!}{
    \begin{tabular}{clccccccccc}
        \toprule
        Estimand & Model & Bias & sd$(\hat\Delta)$ & $\se_{\text{Naive}}(\hat\Delta)$ & $\bbC_{\text{Naive}}(\hat\Delta)$ & $\se_{\text{RVE}}(\hat\Delta)$ & $\bbC_{\text{RVE}}(\hat\Delta)$ & $\se_{\text{RVE}}^{\text{MD}}(\hat\Delta)$ & $\bbC_{\text{RVE}}^{\text{MD}}(\hat\Delta)$ & Convergence \\ 
        \midrule
        \multirow{3}{*}{$\delta(1)$} & EXCH & 0.001 & 0.076 & 0.043 & 76.8 & 0.073 & 93.8 & 0.078 & 95.1 & 100 \\ 
        & NE & 0.000 & 0.075 & 0.064 & 90.8 & 0.072 & 93.8 & 0.077 & 95.4 & 98 \\  
        & DTD & 0.000 & 0.075 & 0.060 & 89.4 & 0.072 & 93.7 & 0.077 & 95.2 & 100 \\ 
        \midrule
        \multirow{3}{*}{$\delta(2)$} & EXCH & 0.001 & 0.102 & 0.056 & 73.5 & 0.099 & 94.9 & 0.107 & 96.3 & 100 \\ 
        & NE & 0.000 & 0.100 & 0.082 & 90.7 & 0.097 & 94.9 & 0.105 & 96.5 & 98 \\ 
        & DTD & -0.000 & 0.099 & 0.086 & 92.4 & 0.096 & 95.0 & 0.104 & 96.3 & 100 \\ 
        \midrule
        \multirow{3}{*}{$\delta(3)$} & EXCH & 0.002 & 0.145 & 0.074 & 70.4 & 0.138 & 93.5 & 0.149 & 95.5 & 100 \\ 
        & NE & 0.002 & 0.144 & 0.105 & 86.1 & 0.136 & 93.5 & 0.147 & 95.5 & 98 \\  
        & DTD & 0.002 & 0.143 & 0.114 & 89.8 & 0.135 & 93.7 & 0.147 & 95.8 & 100 \\ 
        \midrule
        \multirow{3}{*}{$\delta(4)$} & EXCH & 0.002 & 0.197 & 0.100 & 70.2 & 0.184 & 93.2 & 0.202 & 94.8 & 100 \\
        & NE & 0.001 & 0.199 & 0.142 & 85.5 & 0.185 & 93.4 & 0.204 & 95.5 & 98 \\
        & DTD & 0.001 & 0.202 & 0.153 & 87.0 & 0.190 & 93.8 & 0.209 & 95.5 & 100 \\
        \midrule
        \multirow{3}{*}{$\Delta$} & EXCH & 0.001 & 0.119 & 0.060 & 69.7 & 0.113 & 93.5 & 0.123 & 95.2 & 100 \\
        & NE & 0.001 & 0.118 & 0.083 & 85.0 & 0.112 & 94.2 & 0.122 & 95.4 & 98 \\
        & DTD & 0.001 & 0.119 & 0.091 & 87.8 & 0.113 & 93.9 & 0.123 & 95.9 & 100 \\
        \bottomrule
    \end{tabular}
    }\par\vspace{5pt}
    {\scriptsize EXCH: simple exchangeable; NE: nested exchangeable; DTD: discrete-time exponential decay; RVE: robust variance estimator; MD: Mancl and DeRouen.\par}
\end{table}

\begin{table}[htbp]
    \caption{Simulation results based on 2,000 datasets generated under Scenario 2 for Simulation Study VII. The design assumes a standard SW-CRT with $I = 96$, $J = 5$, $\delta = 0$, $\rho_0 = 0.01$, $\rho_1 = 0.1$, $\sigma_\epsilon^{2} = 1$, CAC $= 0.5$, and a continuous period effect $T(t_{ijk}) = \{0.5 t_{ijk}^{2} + \sin(6\pi t_{ijk})\}/J$. For recruitment sizes, we consider (i) balanced recruitment with $K_{ij} = 50$ under both control and intervention, (ii) moderately unbalanced recruitment with $K_{ij} = 25$ under control and $K_{ij} = 75$ under intervention, and (iii) severely unbalanced recruitment with $K_{ij} = 10$ under control and $K_{ij} = 90$ under intervention. For each recruitment size configuration, three recruitment pattern specifications are considered: uniform under both conditions, uniform under control but cluster-period mixed under intervention, and cluster-period mixed under both conditions. $\sd(\cdot)$ denotes the empirical (Monte Carlo) standard deviation. $\se_{\text{Naive}}(\cdot)$, $\se_{\text{RVE}}(\cdot)$, and $\se_{\text{RVE}}^{\text{MD}}(\cdot)$ denote the average model-based, robust, and MD-corrected robust standard errors, respectively. $\bbC_{\text{Naive}}(\cdot)$, $\bbC_{\text{RVE}}(\cdot)$, and $\bbC_{\text{RVE}}^{\text{MD}}(\cdot)$ denote the corresponding empirical coverage of the 95\% confidence intervals.} \label{tab:Unequal96}
    \resizebox{\linewidth}{!}{
    \begin{tabular}{cclccccccccc}
        \toprule
        $K_{ij}$ & Pattern & Model & Bias & sd$(\hat\delta)$ & $\se_{\text{Naive}}(\hat\delta)$ & $\bbC_{\text{Naive}}(\hat\delta)$ & $\se_{\text{RVE}}(\hat\delta)$ & $\bbC_{\text{RVE}}(\hat\delta)$ & $\se_{\text{RVE}}^{\text{MD}}(\hat\delta)$ & $\bbC_{\text{RVE}}^{\text{MD}}(\hat\delta)$ & Convergence \\ 
        \midrule
        \multirow{9}{*}{50 + 50} & \multirow{3}{*}{U + U} & EXCH & 0.000 & 0.043 & 0.023 & 71.3 & 0.041 & 94.1 & 0.042 & 94.8 & 100 \\ 
        & & NE & 0.000 & 0.043 & 0.035 & 88.5 & 0.041 & 93.8 & 0.042 & 94.4 & 98 \\ 
        & & DTD & 0.000 & 0.043 & 0.033 & 86.8 & 0.041 & 93.8 & 0.041 & 94.5 & 100 \\ 
        \cmidrule{2-12}
        & \multirow{3}{*}{U + CP} & EXCH & -0.030 & 0.043 & 0.023 & 62.3 & 0.041 & 88.5 & 0.042 & 89.5 & 100 \\ 
        & & NE & -0.029 & 0.043 & 0.035 & 82.2 & 0.041 & 88.5 & 0.042 & 89.5 & 98 \\ 
        & & DTD & -0.029 & 0.042 & 0.034 & 80.0 & 0.041 & 88.7 & 0.042 & 89.7 & 100 \\ 
        \cmidrule{2-12}
        & \multirow{3}{*}{CP + CP} & EXCH & -0.000 & 0.043 & 0.023 & 72.4 & 0.042 & 94.0 & 0.043 & 94.5 & 100 \\ 
        & & NE & -0.000 & 0.043 & 0.036 & 88.9 & 0.042 & 94.0 & 0.043 & 94.3 & 98 \\
        & & DTD & -0.000 & 0.043 & 0.034 & 88.2 & 0.041 & 94.0 & 0.042 & 94.7 & 100 \\ 
        \midrule
        \multirow{9}{*}{25 + 75} & \multirow{3}{*}{U + U} & EXCH & 0.001 & 0.044 & 0.026 & 75.8 & 0.043 & 94.5 & 0.044 & 95.1 & 100 \\ 
        & & NE & 0.000 & 0.042 & 0.037 & 92.0 & 0.041 & 94.7 & 0.042 & 95.0 & 98 \\ 
        & & DTD & 0.001 & 0.042 & 0.035 & 90.5 & 0.042 & 94.4 & 0.043 & 95.0 & 100 \\ 
        \cmidrule{2-12}
        & \multirow{3}{*}{U + CP} & EXCH & -0.031 & 0.045 & 0.026 & 64.1 & 0.044 & 89.2 & 0.045 & 90.2 & 100 \\ 
        & & NE & -0.032 & 0.043 & 0.038 & 84.0 & 0.042 & 87.8 & 0.043 & 88.3 & 98 \\  
        & & DTD & -0.032 & 0.044 & 0.037 & 82.3 & 0.042 & 87.8 & 0.043 & 88.5 & 100 \\ 
        \cmidrule{2-12}
        & \multirow{3}{*}{CP + CP} & EXCH & -0.004 & 0.044 & 0.026 & 74.9 & 0.044 & 95.3 & 0.045 & 95.7 & 100 \\ 
        & & NE & -0.004 & 0.042 & 0.038 & 92.5 & 0.042 & 95.0 & 0.043 & 95.5 & 98 \\ 
        & & DTD & -0.005 & 0.043 & 0.037 & 91.7 & 0.043 & 95.1 & 0.044 & 95.5 & 100 \\ 
        \midrule
        \multirow{9}{*}{10 + 90} & \multirow{3}{*}{U + U} & EXCH & 0.001 & 0.051 & 0.033 & 80.2 & 0.050 & 94.5 & 0.051 & 95.0 & 100 \\ 
        & & NE & 0.001 & 0.048 & 0.040 & 90.7 & 0.047 & 94.5 & 0.048 & 95.0 & 99 \\ 
        & & DTD & 0.001 & 0.048 & 0.041 & 91.2 & 0.047 & 94.8 & 0.048 & 95.2 & 100 \\ 
        \cmidrule{2-12}
        & \multirow{3}{*}{U + CP} & EXCH & -0.027 & 0.052 & 0.033 & 73.7 & 0.051 & 91.8 & 0.052 & 92.5 & 100 \\ 
        & & NE & -0.029 & 0.048 & 0.043 & 87.7 & 0.047 & 90.3 & 0.048 & 91.0 & 99 \\ 
        & & DTD & -0.030 & 0.049 & 0.043 & 87.5 & 0.048 & 90.8 & 0.049 & 91.3 & 100 \\ 
        \cmidrule{2-12}
        & \multirow{3}{*}{CP + CP} & EXCH & -0.014 & 0.051 & 0.033 & 78.5 & 0.051 & 94.0 & 0.052 & 94.3 & 100 \\ 
        & & NE & -0.014 & 0.048 & 0.043 & 90.8 & 0.047 & 93.4 & 0.048 & 93.9 & 99 \\ 
        & & DTD & -0.015 & 0.048 & 0.043 & 91.5 & 0.048 & 93.5 & 0.049 & 94.2 & 100 \\ 
        \bottomrule
    \end{tabular}
    }\par\vspace{5pt}
    {\scriptsize EXCH: simple exchangeable; NE: nested exchangeable; DTD: discrete-time exponential decay; U: uniform pattern; C: cluster mixed pattern; CP: cluster-period mixed pattern; RVE: robust variance estimator; MD: Mancl and DeRouen.\par}
\end{table}

\begin{table}[htbp]
    \caption{Simulation results based on 2,000 datasets generated under Scenario 2 for Simulation Study VII. The design assumes a standard SW-CRT with $I \in \{16, 32, 96\}$, $J = 5$, $\delta \in \{0, 2\}$, $\rho_0 = 0.01$, $\rho_1 = 0.1$, $\sigma_\epsilon^{2} = 1$, CAC $= 0.5$, and a continuous period effect $T(t_{ijk}) = \{0.5 t_{ijk}^{2} + \sin(6\pi t_{ijk})\}/J$. Two exposure-time-dependent recruitment size configurations are considered: (i) S1 with $K_{ij} = 25$ under control and $K_{ij} = 50, 75, 100, 125$ at exposure times 1, 2, 3, 4, respectively, and (ii) S2 with $K_{ij} = 10$ under control and $K_{ij} = 100, 110, 130, 160$ at exposure times 1, 2, 3, 4, respectively. For each configuration, two recruitment pattern specifications are considered: uniform under both conditions (U + U), and uniform under control but cluster-period mixed under intervention (U + CP). $\sd(\cdot)$ denotes the empirical (Monte Carlo) standard deviation. $\se_{\text{Naive}}(\cdot)$, $\se_{\text{RVE}}(\cdot)$, and $\se_{\text{RVE}}^{\text{MD}}(\cdot)$ denote the average model-based, robust, and MD-corrected robust standard errors, respectively. $\bbC_{\text{Naive}}(\cdot)$, $\bbC_{\text{RVE}}(\cdot)$, and $\bbC_{\text{RVE}}^{\text{MD}}(\cdot)$ denote the corresponding empirical coverage of the 95\% confidence intervals.} \label{tab:unequal_const_16_32_96}
    \resizebox{\linewidth}{!}{
    \begin{tabular}{c|c|cclccccccccc}
        \hline
        $\delta$ & $I$ & Size & Pattern & Model & Bias & sd$(\hat\delta)$ & $\se_{\text{Naive}}(\hat\delta)$ & $\bbC_{\text{Naive}}(\hat\delta)$ & $\se_{\text{RVE}}(\hat\delta)$ & $\bbC_{\text{RVE}}(\hat\delta)$ & $\se_{\text{RVE}}^{\text{MD}}(\hat\delta)$ & $\bbC_{\text{RVE}}^{\text{MD}}(\hat\delta)$ & Convergence\\ 
        \hline
        \multirow{36}{*}{0} & \multirow{12}{*}{16} & \multirow{6}{*}{S1} & \multirow{3}{*}{U + U} & EXCH & -0.004 & 0.110 & 0.066 & 80.2 & 0.098 & 92.9 & 0.111 & 95.8 & 100 \\
        & & & & NE & -0.004 & 0.108 & 0.090 & 91.9 & 0.093 & 92.0 & 0.106 & 95.1 & 100 \\
        & & & & DTD & -0.004 & 0.109 & 0.088 & 90.8 & 0.095 & 91.9 & 0.108 & 95.2 & 100 \\
        \cline{4-14}
        & & & \multirow{3}{*}{U + CP} & EXCH & -0.025 & 0.112 & 0.066 & 78.0 & 0.101 & 92.7 & 0.115 & 95.3 & 100 \\
        & & & & NE & -0.028 & 0.107 & 0.093 & 92.8 & 0.095 & 91.7 & 0.108 & 94.8 & 100 \\
        & & & & DTD & -0.028 & 0.107 & 0.092 & 92.0 & 0.096 & 92.2 & 0.110 & 94.7 & 100 \\
        \cline{3-14}
        & & \multirow{6}{*}{S2} & \multirow{3}{*}{U + U} & EXCH & 0.000 & 0.123 & 0.079 & 83.7 & 0.113 & 92.8 & 0.128 & 95.7 & 100 \\
        & & & & NE & 0.001 & 0.116 & 0.096 & 91.2 & 0.104 & 92.3 & 0.118 & 95.3 & 100 \\
        & & & & DTD & 0.000 & 0.115 & 0.099 & 92.2 & 0.104 & 92.5 & 0.119 & 95.3 & 99 \\
        \cline{4-14}
        & & & \multirow{3}{*}{U + CP} & EXCH & -0.023 & 0.127 & 0.079 & 80.8 & 0.116 & 92.7 & 0.131 & 95.0 & 100 \\ 
        & & & & NE & -0.026 & 0.117 & 0.103 & 92.8 & 0.104 & 92.3 & 0.119 & 95.0 & 100 \\
        & & & & DTD & -0.027 & 0.118 & 0.105 & 93.1 & 0.105 & 92.8 & 0.120 & 95.0 & 100 \\
        \cline{2-14}
        & \multirow{12}{*}{32} & \multirow{6}{*}{S1} & \multirow{3}{*}{U + U} & EXCH & -0.006 & 0.077 & 0.047 & 78.1 & 0.074 & 94.8 & 0.078 & 96.0 & 100 \\ 
        & & & & NE & -0.005 & 0.075 & 0.064 & 92.0 & 0.071 & 94.5 & 0.075 & 95.7 & 99 \\ 
        & & & & DTD & -0.005 & 0.075 & 0.062 & 91.2 & 0.072 & 94.4 & 0.076 & 95.5 & 100 \\ 
        \cline{4-14}
        & & & \multirow{3}{*}{U + CP} & EXCH & -0.027 & 0.080 & 0.047 & 73.7 & 0.075 & 92.4 & 0.080 & 93.8 & 100 \\ 
        & & & & NE & -0.028 & 0.076 & 0.067 & 90.8 & 0.071 & 92.1 & 0.076 & 93.8 & 99 \\ 
        & & & & DTD & -0.028 & 0.076 & 0.065 & 89.8 & 0.072 & 92.5 & 0.077 & 94.2 & 100 \\
        \cline{3-14}
        & & \multirow{6}{*}{S2} & \multirow{3}{*}{U + U} & EXCH & -0.002 & 0.088 & 0.056 & 80.7 & 0.084 & 94.3 & 0.089 & 95.5 & 100 \\  
        & & & & NE & -0.003 & 0.083 & 0.068 & 90.2 & 0.078 & 94.3 & 0.083 & 95.5 & 100 \\  
        & & & & DTD & -0.003 & 0.083 & 0.070 & 91.3 & 0.078 & 94.4 & 0.083 & 95.9 & 100 \\
        \cline{4-14}
        & & & \multirow{3}{*}{U + CP} & EXCH & -0.026 & 0.093 & 0.056 & 76.2 & 0.087 & 93.1 & 0.093 & 94.5 & 100 \\  
        & & & & NE & -0.028 & 0.085 & 0.073 & 90.0 & 0.079 & 92.7 & 0.084 & 93.9 & 100 \\  
        & & & & DTD & -0.028 & 0.085 & 0.074 & 90.0 & 0.080 & 92.8 & 0.085 & 94.1 & 100 \\
        \cline{2-14}
        & \multirow{15}{*}{96} & \multirow{6}{*}{S1} & \multirow{3}{*}{U + U} & EXCH & -0.000 & 0.044 & 0.027 & 78.8 & 0.044 & 94.8 & 0.045 & 95.2 & 100 \\  
        & & & & NE & 0.000 & 0.043 & 0.037 & 92.4 & 0.042 & 95.3 & 0.043 & 95.8 & 98 \\  
        & & & & DTD & 0.000 & 0.043 & 0.036 & 90.5 & 0.043 & 95.2 & 0.044 & 95.5 & 100 \\ 
        \cline{4-14}
        & & & \multirow{3}{*}{U + CP} & EXCH &  -0.025 & 0.047 & 0.027 & 67.9 & 0.045 & 89.7 & 0.046 & 90.7 & 100 \\ 
        & & & & NE & -0.027 & 0.045 & 0.039 & 85.0 & 0.043 & 89.0 & 0.044 & 90.2 & 98 \\ 
        & & & & DTD & -0.027 & 0.045 & 0.038 & 83.2 & 0.043 & 89.6 & 0.044 & 90.7 & 100 \\ 
        \cline{3-14}
        & & \multirow{9}{*}{S2} & \multirow{3}{*}{U + U} & EXCH & 0.000 & 0.049 & 0.032 & 81.3 & 0.050 & 95.5 & 0.051 & 96.0 & 100 \\ 
        & & & & NE & 0.000 & 0.047 & 0.039 & 90.0 & 0.047 & 95.5 & 0.048 & 96.2 & 99 \\
        & & & & DTD & 0.000 & 0.047 & 0.040 & 91.0 & 0.047 & 95.0 & 0.048 & 95.3 & 100 \\ 
        \cline{4-14}
        & & & \multirow{3}{*}{U + CP} & EXCH & -0.026 & 0.053 & 0.032 & 72.0 & 0.052 & 91.7 & 0.053 & 92.2 & 100 \\  
        & & & & NE & -0.028 & 0.048 & 0.042 & 85.9 & 0.047 & 90.5 & 0.048 & 91.0 & 99 \\
        & & & & DTD & -0.029 & 0.049 & 0.043 & 86.3 & 0.047 & 90.0 & 0.048 & 90.6 & 100 \\ 
        \cline{1-1}\cline{4-14}
        \multirow{3}{*}{2} & & & \multirow{3}{*}{U + CP} & EXCH & -0.026 & 0.053 & 0.032 & 72.0 & 0.052 & 91.7 & 0.053 & 92.2 & 100 \\  
        & & & & NE & -0.028 & 0.048 & 0.042 & 85.9 & 0.047 & 90.5 & 0.048 & 91.0 & 97 \\  
        & & & & DTD & -0.029 & 0.049 & 0.043 & 86.3 & 0.047 & 90.0 & 0.048 & 90.6 & 100 \\ 
        \hline
    \end{tabular}
    }\par\vspace{5pt}
    {\scriptsize EXCH: simple exchangeable; NE: nested exchangeable; DTD: discrete-time exponential decay; U: uniform pattern; C: cluster mixed pattern; CP: cluster-period mixed pattern; RVE: robust variance estimator; MD: Mancl and DeRouen.\par}
\end{table}

\begin{table}[htbp]
    \caption{Simulation results based on 2,000 datasets generated under Scenario 4 for Simulation Study VII. The design assumes a standard SW-CRT with $I = 16$, $J = 5$, $\delta(1) \approx -1.3768$, $\delta(2) \approx 0.3831$, $\delta(3) \approx 0.9785$, $\delta(4) \approx 0.0152$, $\Delta = 0$, $\rho_0 = 0.01$, $\rho_1 = 0.1$, $\sigma_\epsilon^{2} = 1$, CAC $= 0.5$, and a continuous period effect $T(t_{ijk}) = \{0.5 t_{ijk}^{2} + \sin(6\pi t_{ijk})\}/J$. Recruitment sizes are set to $K_{ij} = 10$ under control and $K_{ij} = 100, 110, 130, 160$ at exposure times 1, 2, 3, 4, respectively. For each recruitment size configuration, two recruitment pattern specifications are considered: uniform under both conditions (U + U), and uniform under control but cluster-period mixed under intervention (U + CP). $\sd(\cdot)$ denotes the empirical (Monte Carlo) standard deviation. $\se_{\text{Naive}}(\cdot)$, $\se_{\text{RVE}}(\cdot)$, and $\se_{\text{RVE}}^{\text{MD}}(\cdot)$ denote the average model-based, robust, and MD-corrected robust standard errors, respectively. $\bbC_{\text{Naive}}(\cdot)$, $\bbC_{\text{RVE}}(\cdot)$, and $\bbC_{\text{RVE}}^{\text{MD}}(\cdot)$ denote the corresponding empirical coverage of the 95\% confidence intervals.}\label{tab:unequal_TVE_16}
    \resizebox{\linewidth}{!}{
    \begin{tabular}{cclccccccccc}
        \toprule
        Pattern & Estimand & Model & Bias & sd$(\hat\Delta)$ & $\se_{\text{Naive}}(\hat\Delta)$ & $\bbC_{\text{Naive}}(\hat\Delta)$ & $\se_{\text{RVE}}(\hat\Delta)$ & $\bbC_{\text{RVE}}(\hat\Delta)$ & $\se_{\text{RVE}}^{\text{MD}}(\hat\Delta)$ & $\bbC_{\text{RVE}}^{\text{MD}}(\hat\Delta)$ & Convergence \\ 
        \midrule
        & \multirow{3}{*}{$\delta(1)$} & EXCH & 0.001 & 0.130 & 0.100 & 89.7 & 0.118 & 92.8 & 0.134 & 95.4 & 100 \\ 
        & & NE & 0.002 & 0.126 & 0.109 & 92.3 & 0.113 & 92.7 & 0.129 & 95.4 & 98 \\
        & & DTD & 0.001 & 0.126 & 0.108 & 92.2 & 0.113 & 92.4 & 0.129 & 95.7 & 99 \\ 
        \cmidrule{2-12}
        & \multirow{3}{*}{$\delta(2)$} & EXCH &0.004 & 0.180 & 0.134 & 89.2 & 0.161 & 92.6 & 0.185 & 95.4 & 100 \\  
        & & NE & 0.005 & 0.171 & 0.142 & 92.0 & 0.151 & 92.3 & 0.175 & 95.5 & 98 \\  
        & & DTD & 0.004 & 0.169 & 0.148 & 93.8 & 0.149 & 92.0 & 0.173 & 95.3 & 99 \\
        \cmidrule{2-12}
        & \multirow{3}{*}{$\delta(3)$} & EXCH & 0.004 & 0.238 & 0.172 & 88.8 & 0.213 & 92.4 & 0.248 & 95.8 & 100 \\  
        U + U & & NE & 0.005 & 0.229 & 0.180 & 90.5 & 0.201 & 91.6 & 0.235 & 95.1 & 98 \\
        & & DTD & 0.005 & 0.225 & 0.191 & 92.4 & 0.198 & 91.8 & 0.233 & 95.2 & 99 \\
        \cmidrule{2-12}
        & \multirow{3}{*}{$\delta(4)$} & EXCH & 0.004 & 0.306 & 0.215 & 86.5 & 0.270 & 92.2 & 0.319 & 95.5 & 100 \\  
        & & NE & 0.004 & 0.294 & 0.228 & 90.0 & 0.254 & 91.0 & 0.302 & 94.5 & 98 \\ 
        & & DTD & 0.005 & 0.293 & 0.243 & 91.8 & 0.254 & 90.4 & 0.302 & 94.5 & 99 \\ 
        \cmidrule{2-12}
        & \multirow{3}{*}{$\Delta$} & EXCH & 0.003 & 0.205 & 0.151 & 88.7 & 0.183 & 92.0 & 0.212 & 95.5 & 100 \\  
        & & NE & 0.004 & 0.196 & 0.154 & 90.8 & 0.171 & 91.1 & 0.200 & 95.0 & 98 \\ 
        & & DTD & 0.004 & 0.194 & 0.162 & 92.5 & 0.170 & 91.5 & 0.199 & 95.0 & 99 \\  
        \midrule
        & \multirow{3}{*}{$\delta(1)$} & EXCH & -0.029 & 0.130 & 0.101 & 89.4 & 0.120 & 93.0 & 0.136 & 95.8 & 100 \\  
        & & NE & -0.029 & 0.125 & 0.114 & 94.2 & 0.113 & 92.8 & 0.129 & 95.5 & 99 \\
        & & DTD & -0.029 & 0.126 & 0.112 & 94.0 & 0.113 & 92.8 & 0.130 & 95.8 & 100 \\
        \cmidrule{2-12}
        & \multirow{3}{*}{$\delta(2)$} & EXCH & -0.035 & 0.182 & 0.134 & 87.6 & 0.164 & 92.5 & 0.189 & 95.2 & 100 \\ 
        & & NE & -0.036 & 0.171 & 0.146 & 92.4 & 0.152 & 92.0 & 0.176 & 95.0 & 99 \\ 
        & & DTD & -0.037 & 0.169 & 0.154 & 93.5 & 0.150 & 91.6 & 0.175 & 94.8 & 100 \\
        \cmidrule{2-12}
        & \multirow{3}{*}{$\delta(3)$} & EXCH & -0.038 & 0.244 & 0.173 & 86.7 & 0.219 & 92.2 & 0.256 & 95.5 & 100 \\ 
        U + CP & & NE & -0.039 & 0.230 & 0.186 & 91.3 & 0.204 & 91.9 & 0.239 & 95.4 & 99 \\ 
        & & DTD & -0.039 & 0.227 & 0.198 & 93.5 & 0.202 & 92.1 & 0.237 & 95.3 & 100 \\
        \cmidrule{2-12}
        & \multirow{3}{*}{$\delta(4)$} & EXCH & -0.043 & 0.320 & 0.216 & 84.1 & 0.281 & 92.2 & 0.333 & 95.5 & 100 \\ 
        & & NE & -0.044 & 0.303 & 0.237 & 89.8 & 0.260 & 90.3 & 0.312 & 95.2 & 99 \\  
        & & DTD & -0.041 & 0.305 & 0.252 & 91.8 & 0.263 & 90.6 & 0.316 & 95.2 & 100 \\  
        \cmidrule{2-12}
        & \multirow{3}{*}{$\Delta$} & EXCH & -0.036 & 0.209 & 0.152 & 87.0 & 0.187 & 92.8 & 0.217 & 95.5 & 100 \\  
        & & NE & -0.037 & 0.196 & 0.157 & 90.5 & 0.172 & 91.0 & 0.201 & 95.4 & 99 \\  
        & & DTD & -0.037 & 0.195 & 0.167 & 92.3 & 0.172 & 91.3 & 0.202 & 95.4 & 100 \\
        \bottomrule
    \end{tabular}
    }\par\vspace{5pt}
    {\scriptsize EXCH: simple exchangeable; NE: nested exchangeable; DTD: discrete-time exponential decay; U: uniform pattern; CP: cluster-period mixed pattern; RVE: robust variance estimator; MD: Mancl and DeRouen.\par}
\end{table}

\begin{table}[htbp]
    \caption{Simulation results based on 2,000 datasets generated under Scenario 4 for Simulation Study VII. The design assumes a standard SW-CRT with $I = 32$, $J = 5$, $\delta(1) \approx -1.3768$, $\delta(2) \approx 0.3831$, $\delta(3) \approx 0.9785$, $\delta(4) \approx 0.0152$, $\Delta = 0$, $\rho_0 = 0.01$, $\rho_1 = 0.1$, $\sigma_\epsilon^{2} = 1$, CAC $= 0.5$, and a continuous period effect $T(t_{ijk}) = \{0.5 t_{ijk}^{2} + \sin(6\pi t_{ijk})\}/J$. Recruitment sizes are set to $K_{ij} = 10$ under control and $K_{ij} = 100, 110, 130, 160$ at exposure times 1, 2, 3, 4, respectively. For each recruitment size configuration, two recruitment pattern specifications are considered: uniform under both conditions (U + U), and uniform under control but cluster-period mixed under intervention (U + CP). $\sd(\cdot)$ denotes the empirical (Monte Carlo) standard deviation. $\se_{\text{Naive}}(\cdot)$, $\se_{\text{RVE}}(\cdot)$, and $\se_{\text{RVE}}^{\text{MD}}(\cdot)$ denote the average model-based, robust, and MD-corrected robust standard errors, respectively. $\bbC_{\text{Naive}}(\cdot)$, $\bbC_{\text{RVE}}(\cdot)$, and $\bbC_{\text{RVE}}^{\text{MD}}(\cdot)$ denote the corresponding empirical coverage of the 95\% confidence intervals.}\label{tab:unequal_TVE_32}
    \resizebox{\linewidth}{!}{
    \begin{tabular}{cclccccccccc}
        \toprule
        Pattern & Estimand & Model & Bias & sd$(\hat\Delta)$ & $\se_{\text{Naive}}(\hat\Delta)$ & $\bbC_{\text{Naive}}(\hat\Delta)$ & $\se_{\text{RVE}}(\hat\Delta)$ & $\bbC_{\text{RVE}}(\hat\Delta)$ & $\se_{\text{RVE}}^{\text{MD}}(\hat\Delta)$ & $\bbC_{\text{RVE}}^{\text{MD}}(\hat\Delta)$ & Convergence \\  
        \midrule
        & \multirow{3}{*}{$\delta(1)$} & EXCH & -0.003 & 0.092 & 0.072 & 88.8 & 0.088 & 93.5 & 0.093 & 95.2 & 100 \\
        & & NE & -0.003 & 0.089 & 0.077 & 92.1 & 0.085 & 93.7 & 0.091 & 95.3 & 98 \\
        & & DTD & -0.003 & 0.089 & 0.077 & 91.8 & 0.085 & 93.8 & 0.091 & 95.3 & 100 \\ 
        \cmidrule{2-12}
        & \multirow{3}{*}{$\delta(2)$} & EXCH & -0.003 & 0.127 & 0.096 & 87.6 & 0.121 & 93.9 & 0.129 & 95.0 & 100 \\ 
        & & NE & -0.002 & 0.120 & 0.101 & 91.1 & 0.115 & 93.9 & 0.123 & 95.1 & 98 \\ 
        & & DTD & -0.002 & 0.119 & 0.106 & 92.8 & 0.113 & 93.6 & 0.121 & 94.8 & 100 \\ 
        \cmidrule{2-12}
        & \multirow{3}{*}{$\delta(3)$} & EXCH & -0.005 & 0.169 & 0.124 & 86.7 & 0.161 & 93.8 & 0.173 & 95.5 & 100 \\ 
        U + U & & NE & -0.005 & 0.162 & 0.129 & 89.3 & 0.153 & 94.1 & 0.165 & 95.2 & 98 \\  
        & & DTD & -0.004 & 0.160 & 0.137 & 91.5 & 0.151 & 93.8 & 0.163 & 95.5 & 100 \\ 
        \cmidrule{2-12}
        & \multirow{3}{*}{$\delta(4)$} & EXCH & -0.004 & 0.215 & 0.154 & 85.1 & 0.205 & 93.8 & 0.222 & 95.8 & 100 \\  
        & & NE & -0.003 & 0.206 & 0.164 & 88.6 & 0.195 & 93.8 & 0.212 & 95.5 & 98 \\  
        & & DTD & -0.002 & 0.207 & 0.173 & 91.1 & 0.195 & 93.7 & 0.212 & 95.2 & 100 \\ 
        \cmidrule{2-12}
        & \multirow{3}{*}{$\Delta$} & EXCH & -0.004 & 0.145 & 0.108 & 86.8 & 0.138 & 94.0 & 0.148 & 95.3 & 100 \\ 
        & & NE & -0.003 & 0.138 & 0.111 & 89.7 & 0.131 & 93.6 & 0.141 & 95.2 & 98 \\ 
        & & DTD & -0.003 & 0.137 & 0.116 & 91.7 & 0.130 & 93.9 & 0.140 & 95.3 & 100 \\ 
        \midrule
        & \multirow{3}{*}{$\delta(1)$} & EXCH & -0.029 & 0.092 & 0.072 & 86.5 & 0.089 & 92.8 & 0.094 & 94.2 & 100 \\
        & & NE & -0.029 & 0.088 & 0.081 & 92.5 & 0.085 & 92.6 & 0.091 & 94.8 & 98 \\  
        & & DTD & -0.029 & 0.089 & 0.079 & 92.2 & 0.085 & 92.6 & 0.091 & 94.7 & 100 \\
        \cmidrule{2-12}
        & \multirow{3}{*}{$\delta(2)$} & EXCH & -0.031 & 0.126 & 0.096 & 86.5 & 0.123 & 93.8 & 0.131 & 95.5 & 100 \\
        & & NE & -0.031 & 0.118 & 0.104 & 91.8 & 0.114 & 93.3 & 0.123 & 95.5 & 98 \\  
        & & DTD & -0.032 & 0.116 & 0.110 & 93.7 & 0.113 & 92.9 & 0.122 & 95.5 & 100 \\
        \cmidrule{2-12}
        & \multirow{3}{*}{$\delta(3)$} & EXCH &-0.035 & 0.169 & 0.124 & 85.5 & 0.164 & 93.8 & 0.177 & 95.5 & 100 \\
        U + CP & & NE & -0.035 & 0.159 & 0.133 & 90.0 & 0.154 & 93.8 & 0.166 & 95.6 & 98 \\  
        & & DTD & -0.034 & 0.157 & 0.142 & 92.5 & 0.152 & 94.0 & 0.164 & 95.3 & 100 \\
        \cmidrule{2-12}
        & \multirow{3}{*}{$\delta(4)$} & EXCH & -0.034 & 0.219 & 0.155 & 85.0 & 0.213 & 94.3 & 0.231 & 96.4 & 100 \\
        & & NE & -0.032 & 0.206 & 0.170 & 90.3 & 0.199 & 94.2 & 0.217 & 95.8 & 98 \\ 
        & & DTD & -0.028 & 0.207 & 0.180 & 92.0 & 0.201 & 94.2 & 0.219 & 95.7 & 100 \\ 
        \cmidrule{2-12}
        & \multirow{3}{*}{$\Delta$} & EXCH & -0.032 & 0.144 & 0.108 & 86.5 & 0.140 & 94.0 & 0.151 & 95.3 & 100 \\  
        & & NE & -0.032 & 0.134 & 0.113 & 90.5 & 0.130 & 94.0 & 0.141 & 95.5 & 98 \\ 
        & & DTD & -0.031 & 0.134 & 0.119 & 92.3 & 0.130 & 93.8 & 0.140 & 95.3 & 100 \\ 
        \bottomrule
    \end{tabular}
    }\par\vspace{5pt}
    {\scriptsize EXCH: simple exchangeable; NE: nested exchangeable; DTD: discrete-time exponential decay; U: uniform pattern; CP: cluster-period mixed pattern; RVE: robust variance estimator; MD: Mancl and DeRouen.\par}
\end{table}

\begin{table}[htbp]
    \caption{Simulation results based on 2,000 datasets generated under Scenario 4 for Simulation Study VII. The design assumes a standard SW-CRT with $I = 96$, $J = 5$, $\delta(1) \approx -1.3768$, $\delta(2) \approx 0.3831$, $\delta(3) \approx 0.9785$, $\delta(4) \approx 0.0152$, $\Delta = 0$, $\rho_0 = 0.01$, $\rho_1 = 0.1$, $\sigma_\epsilon^{2} = 1$, CAC $= 0.5$, and a continuous period effect $T(t_{ijk}) = \{0.5 t_{ijk}^{2} + \sin(6\pi t_{ijk})\}/J$. Recruitment sizes are set to $K_{ij} = 10$ under control and $K_{ij} = 100, 110, 130, 160$ at exposure times 1, 2, 3, 4, respectively. For each recruitment size configuration, two recruitment pattern specifications are considered: uniform under both conditions (U + U), and uniform under control but cluster-period mixed under intervention (U + CP). $\sd(\cdot)$ denotes the empirical (Monte Carlo) standard deviation. $\se_{\text{Naive}}(\cdot)$, $\se_{\text{RVE}}(\cdot)$, and $\se_{\text{RVE}}^{\text{MD}}(\cdot)$ denote the average model-based, robust, and MD-corrected robust standard errors, respectively. $\bbC_{\text{Naive}}(\cdot)$, $\bbC_{\text{RVE}}(\cdot)$, and $\bbC_{\text{RVE}}^{\text{MD}}(\cdot)$ denote the corresponding empirical coverage of the 95\% confidence intervals.}\label{tab:unequal_TVE_96}
    \resizebox{\linewidth}{!}{
    \begin{tabular}{cclccccccccc}
        \toprule
        Pattern & Estimand & Model & Bias & sd$(\hat\Delta)$ & $\se_{\text{Naive}}(\hat\Delta)$ & $\bbC_{\text{Naive}}(\hat\Delta)$ & $\se_{\text{RVE}}(\hat\Delta)$ & $\bbC_{\text{RVE}}(\hat\Delta)$ & $\se_{\text{RVE}}^{\text{MD}}(\hat\Delta)$ & $\bbC_{\text{RVE}}^{\text{MD}}(\hat\Delta)$ & Convergence \\ 
        \midrule
        & \multirow{3}{*}{$\delta(1)$} & EXCH & -0.000 & 0.053 & 0.041 & 87.5 & 0.052 & 95.0 & 0.053 & 95.2 & 100 \\ 
        & & NE & -0.000 & 0.051 & 0.045 & 91.6 & 0.051 & 94.7 & 0.052 & 95.2 & 96 \\  
        & & DTD & -0.000 & 0.051 & 0.044 & 91.4 & 0.051 & 94.7 & 0.052 & 95.0 & 100 \\ 
        \cmidrule{2-12}
        & \multirow{3}{*}{$\delta(2)$} & EXCH & -0.001 & 0.074 & 0.056 & 87.0 & 0.072 & 94.2 & 0.074 & 94.8 & 100 \\ 
        & & NE & -0.001 & 0.071 & 0.058 & 90.2 & 0.069 & 94.7 & 0.070 & 95.0 & 96 \\ 
        & & DTD & -0.001 & 0.070 & 0.061 & 92.3 & 0.068 & 94.6 & 0.069 & 95.2 & 100 \\ 
        \cmidrule{2-12}
        & \multirow{3}{*}{$\delta(3)$} & EXCH & -0.001 & 0.099 & 0.072 & 85.5 & 0.096 & 93.6 & 0.098 & 94.2 & 100 \\ 
        U + U & & NE & -0.002 & 0.095 & 0.075 & 88.2 & 0.092 & 94.0 & 0.095 & 94.8 & 96 \\  
        & & DTD & -0.001 & 0.094 & 0.079 & 90.7 & 0.091 & 94.8 & 0.093 & 95.1 & 100 \\ 
        \cmidrule{2-12}
        &  \multirow{3}{*}{$\delta(4)$} & EXCH & -0.002 & 0.127 & 0.089 & 84.0 & 0.123 & 94.0 & 0.127 & 94.5 & 100 \\ 
        & & NE & -0.002 & 0.123 & 0.095 & 87.1 & 0.118 & 94.2 & 0.122 & 94.6 & 96 \\ 
        & & DTD & -0.002 & 0.123 & 0.100 & 89.4 & 0.118 & 93.7 & 0.121 & 94.7 & 100 \\  
        \cmidrule{2-12}
        & \multirow{3}{*}{$\Delta$} & EXCH & -0.001 & 0.085 & 0.063 & 86.4 & 0.083 & 94.2 & 0.085 & 94.8 & 100 \\  
        & & NE & -0.001 & 0.082 & 0.064 & 88.6 & 0.079 & 94.3 & 0.081 & 95.0 & 96 \\ 
        & & DTD & -0.001 & 0.081 & 0.067 & 90.2 & 0.078 & 94.6 & 0.080 & 95.1 & 100 \\ 
        \midrule
        & \multirow{3}{*}{$\delta(1)$} & EXCH & -0.031 & 0.053 & 0.041 & 81.7 & 0.053 & 91.2 & 0.054 & 91.5 & 100 \\
        & & NE & -0.031 & 0.052 & 0.047 & 87.7 & 0.051 & 90.7 & 0.052 & 91.2 & 95 \\ 
        & & DTD & -0.031 & 0.052 & 0.046 & 87.0 & 0.051 & 90.7 & 0.052 & 91.2 & 100 \\
        \cmidrule{2-12}
        & \multirow{3}{*}{$\delta(2)$} & EXCH & -0.034 & 0.074 & 0.056 & 82.6 & 0.073 & 91.8 & 0.075 & 92.7 & 100 \\
        & & NE & -0.035 & 0.070 & 0.060 & 87.3 & 0.068 & 91.5 & 0.070 & 92.0 & 95 \\  
        & & DTD & -0.035 & 0.069 & 0.063 & 89.8 & 0.068 & 91.8 & 0.069 & 92.4 & 100 \\ 
        \cmidrule{2-12}
        & \multirow{3}{*}{$\delta(3)$} & EXCH & -0.039 & 0.100 & 0.072 & 81.7 & 0.098 & 92.5 & 0.100 & 93.2 & 100 \\
        U + CP & & NE & -0.038 & 0.095 & 0.077 & 85.8 & 0.092 & 92.4 & 0.095 & 93.3 & 95 \\ 
        & & DTD & -0.037 & 0.094 & 0.082 & 89.3 & 0.091 & 92.4 & 0.094 & 92.9 & 100 \\
        \cmidrule{2-12}
        &  \multirow{3}{*}{$\delta(4)$} & EXCH & -0.042 & 0.129 & 0.090 & 80.9 & 0.127 & 93.5 & 0.131 & 94.0 & 100 \\
        & & NE & -0.041 & 0.123 & 0.098 & 86.2 & 0.120 & 92.8 & 0.124 & 93.3 & 95 \\  
        & & DTD & -0.037 & 0.124 & 0.104 & 88.5 & 0.121 & 92.5 & 0.124 & 93.2 & 100 \\ 
        \cmidrule{2-12}
        & \multirow{3}{*}{$\Delta$} & EXCH & -0.036 & 0.085 & 0.063 & 82.2 & 0.084 & 92.2 & 0.086 & 92.7 & 100 \\ 
        & & NE & -0.036 & 0.080 & 0.066 & 85.6 & 0.078 & 92.0 & 0.080 & 92.7 & 95 \\
        & & DTD & -0.035 & 0.080 & 0.069 & 88.4 & 0.078 & 92.0 & 0.080 & 92.8 & 100 \\  
        \bottomrule
    \end{tabular}
    }\par\vspace{5pt}
    {\scriptsize EXCH: simple exchangeable; NE: nested exchangeable; DTD: discrete-time exponential decay; U: uniform pattern; CP: cluster-period mixed pattern; RVE: robust variance estimator; MD: Mancl and DeRouen.\par}
\end{table}

\begin{table}[htbp]
    \caption{Simulation results based on 2{,}000 datasets for Simulation Study~VIII. We consider two designs, $(I, K_{ij}) = (16, 20)$ and $(32, 50)$, both with $J = 5$, $\delta = 0$, $\rho_0 = 0.01$, $\rho_1 = 0.1$, $\sigma_\epsilon^{2} = 1$, CAC $= 0.5$ under control at a time separation of one period, a continuous period effect $T(t_{ijk}) = \{0.5 t_{ijk}^{2} + \sin(6\pi t_{ijk})\}/J$, a cluster-period mixed recruitment pattern, and a random intervention effect. All working models omit the random intervention effect. The three CTD specifications share the continuous-time decay working correlation and differ in their period effect (Time Effect): discrete period indicators, the oracle basis $\{t_{ijk}^{2}, \sin(6\pi t_{ijk})\}$ matching the true data-generating form, and a natural cubic spline with four degrees of freedom. $\sd(\cdot)$ denotes the empirical (Monte Carlo) standard deviation. $\se_{\text{Naive}}(\cdot)$, $\se_{\text{RVE}}(\cdot)$, and $\se_{\text{RVE}}^{\text{MD}}(\cdot)$ denote the average model-based, robust, and MD-corrected robust standard errors, respectively. $\bbC_{\text{Naive}}(\cdot)$, $\bbC_{\text{RVE}}(\cdot)$, and $\bbC_{\text{RVE}}^{\text{MD}}(\cdot)$ denote the corresponding empirical coverage of the 95\% confidence intervals. Fit time is the average elapsed time in seconds per model fit. CTD fits with $|\hat\delta| > 10$ were excluded before calculating summary measures, leaving 1{,}994--1{,}996 replicates for $I = 32$ and 1{,}998--2{,}000 for $I = 16$ across CTD specifications. The Convergence column reports the package-reported convergence rate among the retained replicates.}\label{tab:ctd_comparison}
    \resizebox{\linewidth}{!}{
    \begin{tabular}{cclccccccccccc}
        \toprule
        $I$ & $K_{ij}$ & Model & Time Effect & Bias & sd$(\hat\delta)$ & $\se_{\text{Naive}}(\hat\delta)$ & $\bbC_{\text{Naive}}(\hat\delta)$ & $\se_{\text{RVE}}(\hat\delta)$ & $\bbC_{\text{RVE}}(\hat\delta)$ & $\se_{\text{RVE}}^{\text{MD}}(\hat\delta)$ & $\bbC_{\text{RVE}}^{\text{MD}}(\hat\delta)$ & Convergence & Fit Time (s) \\
        \midrule
        \multirow{8}{*}{16} & \multirow{8}{*}{20}
          & EXCH & Discrete             & -0.001 & 0.120 & 0.087 & 87.8 & 0.108 & 93.0 & 0.124 & 96.3 & 100.0 & $<0.1$ \\
        & & NE   & Discrete             & -0.001 & 0.120 & 0.104 & 93.4 & 0.108 & 92.5 & 0.123 & 96.0 & 99.7  & $<0.1$ \\
        & & DTD  & Discrete             & -0.001 & 0.121 & 0.105 & 93.8 & 0.108 & 93.2 & 0.123 & 95.8 & 99.7  & $<1$ \\
        & & CTD  & Discrete             &  0.003 & 0.138 & 0.106 & 92.5 & --    & --   & --    & --   & 100.0 & 31 \\
        & & CTD  & Continuous (oracle)  &  0.001 & 0.135 & 0.099 & 88.7 & --    & --   & --    & --   & 100.0 & 37 \\
        & & CTD  & Continuous (spline)  &  0.024 & 0.136 & 0.099 & 89.7 & --    & --   & --    & --   & 100.0 & 32 \\
        \midrule
        \multirow{8}{*}{32} & \multirow{8}{*}{50}
          & EXCH & Discrete             &  0.000 & 0.071 & 0.040 & 75.5 & 0.070 & 94.2 & 0.074 & 95.7 & 100.0 & $<1$ \\
        & & NE   & Discrete             &  0.000 & 0.071 & 0.061 & 91.2 & 0.069 & 94.0 & 0.074 & 95.8 & 99.0  & $<1$ \\
        & & DTD  & Discrete             &  0.000 & 0.071 & 0.059 & 90.7 & 0.069 & 93.9 & 0.074 & 95.7 & 100.0 & $<1$ \\
        & & CTD  & Discrete             &  0.001 & 0.073 & 0.064 & 90.1 & --    & --   & --    & --   & 100.0 & 3626 \\
        & & CTD  & Continuous (oracle)  & -0.005 & 0.199 & 0.057 & 84.8 & --    & --   & --    & --   & 100.0 & 2662 \\
        & & CTD  & Continuous (spline)  &  0.039 & 0.077 & 0.055 & 79.8 & --    & --   & --    & --   & 100.0 & 2621 \\
        \bottomrule
    \end{tabular}
    }\par\vspace{5pt}
    {\scriptsize EXCH: simple exchangeable; NE: nested exchangeable; DTD: discrete-time exponential decay; CTD: continuous-time decay; RVE: robust variance estimator; MD: Mancl and DeRouen. Time Effect specifications for the CTD rows: Discrete (period indicators), continuous oracle (true basis $t_{ijk}^2$ and $\sin(6\pi t_{ijk})$), and continuous spline (natural cubic, degree of freedom $=4$).\par}
\end{table}

\begin{table}[t]
\caption{Batched stepped wedge design of the Australian Disinvestment Trial. Each hospital forms a batch of six wards followed over seven monthly periods. Wards are grouped by hospital and ordered by crossover time within each hospital. Cluster identifiers follow the analysis dataset. A value of 0 indicates the existing weekend service, 1 indicates service removal, and $\cdot$ indicates a calendar period outside the scheduled follow-up for that batch.}\label{table:design_matrix}
\centering
    \begin{tabular}{ccccccccccc}
        \toprule
        \multirow{2}{*}{Hospital (batch)} & \multirow{2}{*}{Cluster} & \multicolumn{9}{c}{Calendar period} \\
        \cmidrule(lr){3-11}
         & & 1 & 2 & 3 & 4 & 5 & 6 & 7 & 8 & 9 \\
        \midrule
        \multirow{6}{*}{1} & 5 & 0 & 1 & 1 & 1 & 1 & 1 & 1 & $\cdot$ & $\cdot$ \\
         & 1 & 0 & 0 & 1 & 1 & 1 & 1 & 1 & $\cdot$ & $\cdot$ \\
         & 6 & 0 & 0 & 0 & 1 & 1 & 1 & 1 & $\cdot$ & $\cdot$ \\
         & 2 & 0 & 0 & 0 & 0 & 1 & 1 & 1 & $\cdot$ & $\cdot$ \\
         & 4 & 0 & 0 & 0 & 0 & 0 & 1 & 1 & $\cdot$ & $\cdot$ \\
         & 3 & 0 & 0 & 0 & 0 & 0 & 0 & 1 & $\cdot$ & $\cdot$ \\
        \midrule
        \multirow{6}{*}{2} & 7 & $\cdot$ & $\cdot$ & 0 & 1 & 1 & 1 & 1 & 1 & 1 \\
         & 8 & $\cdot$ & $\cdot$ & 0 & 0 & 1 & 1 & 1 & 1 & 1 \\
         & 9 & $\cdot$ & $\cdot$ & 0 & 0 & 0 & 1 & 1 & 1 & 1 \\
         & 10 & $\cdot$ & $\cdot$ & 0 & 0 & 0 & 0 & 1 & 1 & 1 \\
         & 11 & $\cdot$ & $\cdot$ & 0 & 0 & 0 & 0 & 0 & 1 & 1 \\
         & 12 & $\cdot$ & $\cdot$ & 0 & 0 & 0 & 0 & 0 & 0 & 1 \\
        \bottomrule
    \end{tabular}
\end{table}

\begin{table}[!htbp]
\caption{Sensitivity of estimated intervention effects (in days), standard errors, and 95\% CIs to independent and correlated RI effects under EXCH and NE for the Australian Disinvestment Trial. All models include hospital-specific period effects. The cross indicates a constant intervention effect, and the check indicates an exposure-time-dependent intervention effect.}\label{table:empirical_RI}
\resizebox{\linewidth}{!}{
    \begin{tabular}{cllclcc}
        \toprule
        \textbf{Time Varying} & \textbf{Working Model} & \textbf{RI Specification} & \textbf{Estimate} & \textbf{Variance Estimator} & \textbf{Estimated Standard Error} & \textbf{95\% CI} \\
        \midrule
        \multirow{12}{*}{$\newcrossmark$} & \multirow{6}{*}{EXCH} & \multirow{3}{*}{Independent} & \multirow{3}{*}{$1.63$} & Model-based & $0.29$ & $(0.98, 2.28)$ \\
         &  &  &  & RVE & $0.31$ & $(0.94, 2.32)$ \\
         &  &  &  & RVE + MD & $0.41$ & $(0.71, 2.55)$ \\
        \cmidrule{3-7}
         &  & \multirow{3}{*}{Correlated} & \multirow{3}{*}{$1.63$} & Model-based & $0.30$ & $(0.96, 2.30)$ \\
         &  &  &  & RVE & $0.31$ & $(0.94, 2.32)$ \\
         &  &  &  & RVE + MD & $0.41$ & $(0.71, 2.55)$ \\
        \cmidrule{2-7}
         & \multirow{6}{*}{NE} & \multirow{3}{*}{Independent$^\dagger$} & \multirow{3}{*}{$1.70$} & Model-based & $0.38$ & $(0.85, 2.55)$ \\
         &  &  &  & RVE & $0.33$ & $(0.97, 2.43)$ \\
         &  &  &  & RVE + MD & $0.44$ & $(0.71, 2.69)$ \\
        \cmidrule{3-7}
         &  & \multirow{3}{*}{Correlated$^\dagger$} & \multirow{3}{*}{$1.71$} & Model-based & $0.38$ & $(0.86, 2.56)$ \\
         &  &  &  & RVE & $0.32$ & $(0.98, 2.43)$ \\
         &  &  &  & RVE + MD & $0.44$ & $(0.72, 2.69)$ \\
        \midrule
        \multirow{12}{*}{$\newcheckmark$} & \multirow{6}{*}{EXCH} & \multirow{3}{*}{Independent} & \multirow{3}{*}{$2.64$} & Model-based & $0.44$ & $(1.65, 3.63)$ \\
         &  &  &  & RVE & $0.39$ & $(1.77, 3.52)$ \\
         &  &  &  & RVE + MD & $0.55$ & $(1.41, 3.87)$ \\
        \cmidrule{3-7}
         &  & \multirow{3}{*}{Correlated} & \multirow{3}{*}{$2.56$} & Model-based & $0.43$ & $(1.60, 3.52)$ \\
         &  &  &  & RVE & $0.38$ & $(1.72, 3.41)$ \\
         &  &  &  & RVE + MD & $0.51$ & $(1.43, 3.69)$ \\
        \cmidrule{2-7}
         & \multirow{6}{*}{NE} & \multirow{3}{*}{Independent$^\dagger$} & \multirow{3}{*}{$2.58$} & Model-based & $0.58$ & $(1.30, 3.86)$ \\
         &  &  &  & RVE & $0.39$ & $(1.71, 3.45)$ \\
         &  &  &  & RVE + MD & $0.49$ & $(1.49, 3.67)$ \\
        \cmidrule{3-7}
         &  & \multirow{3}{*}{Correlated$^\dagger$} & \multirow{3}{*}{$2.56$} & Model-based & $0.56$ & $(1.30, 3.81)$ \\
         &  &  &  & RVE & $0.39$ & $(1.69, 3.43)$ \\
         &  &  &  & RVE + MD & $0.49$ & $(1.46, 3.65)$ \\
        \bottomrule
    \end{tabular}
    }\par\vspace{5pt}
    {\scriptsize EXCH: simple exchangeable; NE: nested exchangeable; DTD: discrete-time exponential decay; RVE: robust variance estimator; MD: Mancl and DeRouen; CI: confidence interval. $^\dagger$A boundary (singular) fit was identified.\par}
\end{table}